\documentclass[prd,superscriptsize,reprint]{revtex4-1}
\usepackage[utf8]{inputenc}

\usepackage{graphicx,xcolor,overpic,mathtools}
\usepackage{amsthm,amsmath,amssymb}
\usepackage[colorlinks]{hyperref}
\usepackage{textcomp}
\definecolor{coolblack}{rgb}{0.0, 0.18, 0.39}
\usepackage{orcidlink}
\hypersetup{
    colorlinks = false,
}
\PassOptionsToPackage{normalem}{ulem}
\usepackage{ulem}
\usepackage{bm}
\usepackage{url}
\usepackage{physics}
\usepackage[makeroom]{cancel}
\usepackage{cleveref}
\usepackage{tensor}
\usepackage{mathrsfs}
\usepackage{slashed}
\usepackage{caption}
\usepackage[justification=justified,singlelinecheck=false]{caption}

\makeatletter
\long\def\@makecaption#1#2{%
  \vskip\abovecaptionskip
  \begingroup\small
    \spaceskip=0.30em plus 0.05em minus 0.10em   % tight interword space
    \emergencystretch=0.5em
    \hyphenpenalty=10
    \frenchspacing
    \setlength{\leftskip}{0pt}\setlength{\rightskip}{0pt}%
    \noindent #1: #2\par
  \endgroup
  \vskip\belowcaptionskip}
\makeatother

\usepackage{subcaption}
\usepackage[mode=buildnew]{standalone}

\newcommand{\comment}[1]{}

\usepackage{pgfplots}
\usepackage{diagbox}
\usepackage{stackengine,amssymb,graphicx}

\NewDocumentCommand{\evat}{sO{\bigg}mm}{%
  \IfBooleanTF{#1}
   {\mleft. #3 \mright|_{#4}}
   {#3#2|_{#4}}%
}
\usepackage{enumerate}
\definecolor{azure}{rgb}{0.0, 0.5, 1.0}

\begin{document}

\title[]{Scalar-Assisted Growth of Supermassive Black Holes: Boson-Mass Windows and Observational Constraints}

\author{Jorge Castelo Mourelle\orcidlink{0000-0002-0074-2608}} 
\email{jorge.castelo@correo.nucleares.unam.mx}
\affiliation{Instituto de Ciencias Nucleares, Universidad Nacional Autónoma de México, Circuito Exterior C.U., A.P. 70-543, México D.F. 04510, México.}

\author{Darío Núñez\orcidlink{0000-0002-3235-4562}} 
\email{nunez@nucleares.unam.mx}
\affiliation{Instituto de Ciencias Nucleares, Universidad Nacional Autónoma de México, Circuito Exterior C.U., A.P. 70-543, México D.F. 04510, México.}

\author{Saeed Fakhry\orcidlink{0000-0002-6349-8489}} 
\email{saeed.fakhry@uv.es}
\affiliation{Departamento de Astronomía y Astrofísica, Universitat de València, Avinguda Vicent 
Andrés Estellés 19, 46100, Burjassot (València), Spain}

\author{Nicolas Sanchis-Gual} 
\email{nicolas.sanchis@uv.es}
\affiliation{Departamento de Astronomía y Astrofísica, Universitat de València, Avinguda Vicent 
Andrés Estellés 19, 46100, Burjassot (València), Spain}

\date{\today}

\begin{abstract}
We study a phenomenological cosmological model for black-hole growth in which a finite baryonic reservoir and a bosonic scalar cloud feed the black hole simultaneously. The baryonic component is accreted through an Eddington-like channel, while the scalar cloud transfers mass through an effective quasibound-state decay rate. Evolving the system directly in redshift within a flat $\Lambda$CDM background, we perform a broad scan over $\mu$, $M_0$, $M_{c,0}$, and $M_{b,0}$ to determine whether scalar-assisted accretion can accelerate the formation of $10^9M_\odot$ supermassive black holes. We find that the scalar contribution is highly selective in parameter space. For the black hole mass scales relevant to the present supermassive-black-hole-growth problem, scalar transfer becomes particularly efficient for $\mu\sim10^{-19}$-$10^{-16}\,\mathrm{eV}$, triggering rapid cloud depletion and substantially earlier black hole growth, whereas outside this region the evolution remains predominantly baryonic. More generally, this efficient window reflects a preferred range of the dimensionless coupling $\alpha=\mu M_{\rm BH}$ and therefore shifts approximately as $\mu\propto M_{\rm BH}^{-1}$ with the characteristic black hole mass. We further compare the model with observed quasar masses and luminosities, finding that both diagnostics favor lower boson masses, broadly around $\mu\sim10^{-21}$--$10^{-20}\,\mathrm{eV}$. These two results need not be interpreted as contradictory: the boson-mass scale that most efficiently accelerates very early black hole growth depends on the characteristic black hole mass and evolutionary stage, while a lighter, fuzzy-dark-matter-like scalar can remain compatible with the observed quasar population at later epochs. This separation between dynamical efficiency and observational compatibility leaves open the possibility that different bosonic mass scales may play complementary roles in the growth of supermassive black holes and dark-matter phenomenology.
\end{abstract}

\maketitle

% --
\section{Introduction}\label{sec:intro}
% --

 The existence of supermassive black holes (SMBHs) at very early cosmic
epochs remains one of the most compelling open problems in modern
astrophysics and cosmology. Wide-area quasar surveys have consistently uncovered a population of hyperluminous active galactic nuclei powered by SMBHs with masses in the range $10^{9}\text{-}10^{10}\,M_{\odot}$ at redshifts $z\gtrsim 6$~\cite{2006AJ....131.1203F,2011Natur.474..616M, 2015Natur.518..512W,Banados2018,2018Natur.553..473B,Yang2020, 2021ApJ...907L...1W,2021ApJ...908L..33Y,Wang2021}. This picture has been reinforced by the James Webb Space Telescope (JWST), which is now uncovering an abundant population of active galactic nuclei and massive black holes (BH) candidates at $z\gtrsim 8$, including the accreting BH in GN-z11 at $z=10.6$ and X-ray luminous sources such as UHZ-1 at $z\simeq 10.1$~\cite{harikane2023jwst,Maiolino2024,Bogdan2024, Goulding2023}. More recently, JWST has also revealed the so-called Little Red Dots (LRDs), a population of compact, red high-redshift sources, many of which exhibit broad Balmer emission lines and other signatures consistent with accreting BHs~\cite{2024ApJ...964...39G,2024ApJ...963..129M}. These objects appear remarkably abundant at $z\sim4-7$ and may trace an important phase of early BH growth, although their exact physical nature remains under active debate. Recent spectroscopic and dynamical studies further support the presence of BHs with masses in the range $\sim10^{5}-10^{7}M_\odot$ in at least part of the LRD population~\cite{2026Natur.653.1017J}. The extreme youth of these systems, less than about $800\,\mathrm{Myr}$ after the Big Bang, places stringent constraints on how much mass a BH seed can acquire within the available cosmic time, and provides the strongest motivation to investigate alternative or complementary channels of early BH growth~\cite{Volonteri2010,Inayoshi2020,2020ARA&A..58...27I}.
Several theoretical mechanisms have been developed to account for this rapid assembly. Remnants of Population III stars can provide light seeds of $10^{1}\text{-}10^{3}\,M_{\odot}$~\cite{2001ApJ...551L..27M, 2011Sci...334..948T,2010ARA&A..48..339B,2004ApJ...604..484M}, while the direct collapse of pristine gas clouds or the formation of supermassive stars can yield heavier seeds of $10^{4}\text{-}10^{6}\,M_{\odot}$~\cite{1994ApJ...432...52L, 2006MNRAS.370..289B,2010MNRAS.402.1249S,2017MNRAS.469.3329W, Shlosman:2015wma,Haemmerle2018,Woods2019}. Runaway stellar collisions in dense clusters offer a further route to heavy seeds~\cite{2004Natur.428..724P,Gaete2024}. Subsequent growth may proceed through sustained Eddington-limited accretion~\cite{Salpeter1964,Volonteri2010}, super-Eddington episodes~\cite{2006MNRAS.370..289B,Trinca2024}, and mergers in dense stellar or galactic environments~\cite{Atallah2023,Gaete2024}. Under favorable circumstances, these channels can generate very massive BHs at high redshift, but their efficiency depends sensitively on the seed population, radiative efficiency, duty cycle, gas supply, and merger history~\cite{Volonteri2010,Inayoshi2020,2020ARA&A..58...27I}.

The difficulty is not simply the availability of baryonic matter, but the rate at which it can be converted into BH mass. For radiatively efficient accretion, the Eddington luminosity sets the characteristic scale for steady inflow: as the accretion rate increases, radiation pressure acts back on the surrounding gas and limits the net mass supply to the BH~\cite{Salpeter1964,Volonteri2010,Inayoshi2020}. Sustained Eddington growth therefore proceeds on the Salpeter timescale and requires a sufficiently massive seed, a high duty cycle, and a prolonged gas supply to produce $\sim10^9M_\odot$ BHs by $z\gtrsim6$--$10$~\cite{Volonteri2010,Inayoshi2020}. Super-Eddington accretion, massive initial seeds, repeated mergers, and dense nuclear environments can alleviate these requirements~\cite{2006MNRAS.370..289B,Trinca2024,Atallah2023,Gaete2024}, but each relies on specific astrophysical conditions. This motivates the exploration of additional channels that can transfer mass to the BH without being subject to the same radiative feedback that regulates ordinary baryonic accretion.

Dark matter provides a natural setting in which to search for such mechanisms. Ultralight bosonic fields are well motivated in theories containing axions, axion-like particles, and string-theory moduli~\cite{1977PhRvL..38.1440P,1978PhRvL..40..223W,
1978PhRvL..40..279Wi,2006JHEP...06..051S,2010PhRvD..81l3530A}, and have attracted sustained interest as viable dark-matter candidates on both galactic and cosmological scales~\cite{Matos:1998vk,Matos:2004rs, Hu2000,Schive2014,Hui2017,Niemeyer2020,Pozo2025,Mourelle:2024dlt,Mourelle:2025ilv,Pombo:2025xsv,bernal2025natural}. On astrophysical scales, they can form coherent, quasi-macroscopic configurations such as solitonic cores, boson stars, and gravitational atoms~\cite{Liebling:2012fv,sanchis2021multifield,Herdeiro:2014goa,
Herdeiro:2015waf,guzman2020gravitational,guzman2022possible,Sanchis-Gual:2020mzb,Gomez:2026xru}. Dense scalar configurations may enhance the inflow of baryonic gas onto a BH~\cite{Chiu2025}, while direct capture or absorption of scalar matter provides an additional source of BH mass~\cite{Barranco2011,Barranco2012,Barranco2014,Cardoso2022, Annulli:2020lyc,Imai:2025ate}. Primordial black holes also offer a further
dark-matter-inspired route to early massive
objects~\cite{2020ARNPS..70..355C}.

A massive bosonic field around a BH generically develops \emph{quasibound states}: spatially localized configurations satisfying purely ingoing conditions at the horizon and decaying at
infinity, whose frequencies are complex \cite{Detweiler1980,2007PhRvD..76h4001D,Barranco2012,Brito15}. Depending on the geometry, these states either decay through absorption by the horizon or, for suitable modes around a rotating BH, grow through the classical superradiant instability~\cite{1971JETPL..14..180Z,1972JETP...35.1085Z, 1973JETP...37...28S,1972Natur.238..211P,Detweiler1980, 1979AnPhy.118..139Z,2007PhRvD..76h4001D,2013PhRvD..87l4026D,Brito15}. Superradiance can extract a substantial fraction of the BH rotational energy, form long-lived scalar clouds, and give rise to distinctive
signatures in gravitational-wave and BH-spin observations~\cite{2011PhRvD..83d4026A,2014PTEP.2014d3E02Y, 2012PhRvD..86f4036W,2018PhRvL.121m1104E,2012PhRvD..86f4019P,
2017PhRvD..96c5019B,2017PhRvL.119d1101E,2021PhRvD.103i5019B,
2019JCAP...12..006B,2017PhRvL.119m1101B,2017PhRvD..96f4050B,
2019PhRvL.123q1101P,2021PhRvL.126o1102N,2019NatAs...3..447H,
2023PhRvD.107j4003S,2021NatAs...5..133R,2021SSRv..217...65B,
2023PhRxX..13d1039A}. Extensive work on hairy BH solutions and their long-term (in)stability further characterizes the endpoint of this process~\cite{sanchis2016explosion,
bosch2016nonlinear,Degollado:2018ypf,ganchev2018scalar,nicoules2026splitting}.
For the decaying, spherically symmetric quasibound states most relevant to the cosmological setting considered here, the imaginary part of the frequency depends strongly on the dimensionless combination $\mu M_{\rm BH}$, where $\mu$ is the boson mass and $M_{\rm BH}$ is the BH mass~\cite{Detweiler1980,1976PhRvD..14.3251U, 1976PhRvD..13..198P,2005PhLB..609..377K,Barranco2011,Barranco2012, Barranco2014}. In the small-coupling limit, the decay rate of the dominant scalar mode scales as $\Gamma\propto\mu^{6} M_{\rm BH}^{5}$, so that a cloud initially absorbed only very slowly may enter a much faster depletion phase as the BH grows. This strong sensitivity to $\mu M_{\rm BH}$ is the physical origin of the nonlinear feedback between the scalar reservoir and the growing horizon. Numerical evolutions of massive scalars on accreting or dynamical BH backgrounds~\cite{SanchisGual2015a, SanchisGual2016,east2014black, deCesare:2026fie}, together with fully relativistic simulations of a BH embedded in a self-gravitating bosonic soliton~\cite{Cardoso2022}, consistently support the existence of rapid, catastrophic absorption regimes when the coupling $\mu M_{\rm BH}$ becomes favorable. %, in sharp contrast with the naive slow-absorption expectation.

Building on these results, Ref.~\cite{SanchisGual2026DarkToBlack} introduced an effective ``dark-to-black'' mechanism in which the absorption of a finite scalar reservoir increases the BH mass and thereby accelerates the subsequent quasibound-state decay. That work established that the scalar channel alone can undergo a strong positive-feedback phase and transfer a substantial fraction of the surrounding bosonic matter to the BH on cosmologically relevant timescales. The present work addresses a different question: whether this mechanism remains astrophysically relevant once ordinary baryonic growth is included simultaneously. To this end, we evolve for the first time within this framework a BH coupled to two finite reservoirs-baryonic and bosonic-in a cosmological $\Lambda$CDM background, and follow their competition throughout the full redshift evolution. Rather than asking only whether scalar absorption can be rapid, we quantify when it materially changes the BH growth history relative to an Eddington-like baryonic control, how this depends jointly on $\mu$, $M_0$, $M_{c,0}$, and $M_{b,0}$, and which regions of parameter space become scalar- or baryon-dominated. We further confront the resulting trajectories with observed quasar masses and luminosities. This allows us to distinguish the boson mass regime that maximizes scalar-assisted growth from the regime that best matches observations.

In this work, we investigate, within a controlled phenomenological framework, whether an ultralight bosonic reservoir can provide a complementary channel for BH growth and contribute to the early assembly of SMBHs, or whether its impact remains confined to restricted regions of parameter space. This question is particularly timely given the growing population of massive BHs inferred at $z\gtrsim7$--$10$ from high-redshift quasars and JWST observations~\cite{Banados2018,Yang2020,Wang2021,Maiolino2024,Bogdan2024,harikane2023jwst,Goulding2023}, together with increasingly stringent constraints on ultralight bosonic fields from small-scale structure, BH spins, and continuous gravitational-wave searches~\cite{Hui2017,Niemeyer2020,Pozo2025,2019PhRvL.123q1101P,2021PhRvL.126o1102N,2019NatAs...3..447H,2023PhRvD.107j4003S,2021NatAs...5..133R}. We therefore quantify when scalar-assisted growth becomes genuinely competitive with baryonic accretion, how the corresponding efficient boson-mass window depends on the BH mass scale through the combination $\mu M_{\rm BH}$, and whether the resulting growth histories remain compatible with the observed high-redshift quasar population. In this way, the model connects the microscopic scalar-transfer process with cosmological BH assembly and with complementary observational diagnostics based on BH masses and luminosities.

The structure of this work is as follows. In Sec.~\ref{sec:setup}, we introduce the theoretical setup, including the flat $\Lambda$CDM background, the dynamical variables associated with the BH and the two finite matter reservoirs, and the effective prescriptions adopted for the baryonic Eddington-like accretion channel and the bosonic quasibound-state transfer channel. In Sec.~\ref{sec:numerics}, we provide the numerical implementation used to integrate the coupled system directly in redshift, together with the parameter scan spanning the boson mass $\mu$, the initial BH seed mass $M_0$, the initial scalar-cloud mass $M_{c,0}$, and the initial baryonic reservoir mass $M_{b,0}$. In Sec.~\ref{sec:results}, we present illustrative redshift evolutions of the three mass components, followed by the analysis of the arrival redshift $z_{\rm reach}(10^{9}M_\odot)$ and the full multidimensional parametric study of the relative importance of the scalar and baryonic accretion channels. We also discuss the preliminary comparison with the observational quasar sample and the statistical ranking of the explored models. Finally, Sec.~\ref{sec:conclusions}, we summarize the main conclusions, discusses the limitations of the present phenomenological framework, and outlines possible extensions toward a more complete relativistic treatment. We also added two insightful appendices: \ref{app:clouds}, which treats the possibility of scalar clouds at high redshifts, and \ref{app:trajplots}, which complements the analysis.

% ---
\section{Theoretical setup}
\label{sec:setup}
% ---

The theoretical framework adopted in this work describes the cosmological growth of a single BH embedded simultaneously in two finite-mass reservoirs: a bosonic scalar cloud, associated with an ultralight dark-matter component that surrounds the BH, and a baryonic component, representing the ordinary matter available for accretion in the neighborhood of the BH. The purpose of the construction is not to reproduce a fully self-consistent relativistic bosonic configuration in the strong-gravity regime, nor to model the detailed hydrodynamics of the baryonic environment. Rather, the goal is to provide an effective and controlled framework in which the relative importance of the scalar-assisted and baryonic-accretion channels can be systematically explored across a broad multidimensional parameter space, while retaining the physical essentials of both mechanisms. In this sense, the model is phenomenological in nature: each transfer channel is represented by an effective rate that captures the dominant scaling identified in the corresponding microscopic analyses, and the free parameters of the two reservoirs are treated as initial conditions to be scanned.

The cosmological evolution is followed on a spatially flat $\Lambda$CDM background, in which the Hubble rate as a function of redshift $z$ reads
\begin{equation}
H(z)=H_{0}\sqrt{\Omega_{r}(1+z)^{4}+\Omega_{m}(1+z)^{3}
+\Omega_{\Lambda}},
\label{eq:Hz}
\end{equation}
where $H_{0}\equiv H(z=0)$ is the present-day Hubble constant, and $\Omega_{r}$, $\Omega_{m}$, and $\Omega_{\Lambda}$ denote the current fractional energy densities in radiation, non-relativistic matter, and cosmological constant, respectively. The relation between the
redshift and the scale factor $a(t)$ is
\begin{equation}
1+z=\frac{a_{0}}{a(t)},
\label{eq:aofz}
\end{equation}
with $a_{0}=1$ at the present epoch, so that larger values of $z$
correspond to earlier cosmic times, while $z=0$ denotes the present
Universe. The differential map between cosmic time and redshift then
takes the standard form
\begin{equation}
\frac{dt}{dz}=-\frac{1}{(1+z)\,H(z)}, \label{eq:dtdz} \end{equation} 
which fixes the sign convention used throughout the integration: because the evolution is followed from high to low redshift, BH growth translates into a negative derivative $dM_{\rm BH}/dz$, and the depletion of the two reservoirs corresponds to positive derivatives of their masses with respect to $z$. This convention is convenient for the numerical implementation but has no physical
consequence.

The dynamical state of the system is fully specified by the
three-dimensional vector
\begin{equation}
Y(z)=\bigl(M_{\rm BH}(z),\,M_{c}(z),\,M_{b}(z)\bigr),
\label{eq:Yvec}
\end{equation}
where $M_{\rm BH}$ is the mass of the growing BH, $M_{c}$ is the
mass remaining in the scalar cloud, and $M_{b}$ is the mass still
available in the baryonic reservoir. Both $M_{c}$ and $M_{b}$ are
treated as finite: they act as sources that feed the BH through two distinct effective channels, but they are progressively depleted as the evolution proceeds. This finite-reservoir prescription is a key feature of the framework, as it enforces mass conservation between the BH and its surrounding environment and, more importantly, allows the model to capture the essential competition between rapid initial absorption and eventual exhaustion of a given matter component.

The baryonic contribution is modeled through a standard Eddington-like accretion prescription \cite{Salpeter1964, Inayoshi2020}. Assuming a radiative efficiency $\epsilon$ and defining the Eddington timescale $t_{\rm Edd}$, the maximal baryonic mass rate feeding the BH reads
\begin{equation}
\dot M_{\rm edd}
=\frac{1-\epsilon}{\epsilon}\,\frac{M_{\rm BH}}{t_{\rm Edd}},
\label{eq:medd}
\end{equation}
which corresponds to the classical exponential Salpeter law with an $e$-folding time $t_{\rm Sal}=\epsilon\,t_{\rm Edd}/(1-\epsilon)$. In the reference numerical implementation, the fiducial values $\epsilon=0.1$ and $t_{\rm Edd}\simeq 4.5\times 10^{8}\,{\rm yr}$ are adopted, in agreement with the standard thin-disk assumptions commonly employed in high-redshift SMBH growth studies \cite{Volonteri2010, Inayoshi2020}. The baryonic channel is active only as long as baryonic material remains in the reservoir. The effective baryonic mass transfer rate is therefore expressed as
\begin{equation}
\dot M_{\rm b,acc} =
\begin{cases}
\dot M_{\rm edd}, & M_{b}>0,\\[2mm]
0, & M_{b}=0,
\end{cases}
\label{eq:mbacc}
\end{equation}
so that the BH cannot accrete more baryonic mass than is contained
in the surrounding reservoir. This limited-reservoir constraint marks a substantive departure from the idealized picture of an infinite gas supply and allows the model to explore self-consistently the transition between a baryonically driven regime and a baryon-exhausted regime.

The bosonic contribution, in turn, is described by an effective scalar growth rate that reflects the small-coupling scaling of the dominant quasibound-state decay mode of a scalar field in the vicinity of a Schwarzschild-like BH. In this regime, the imaginary part of the frequency of the dominant $\ell=0$ mode scales as $\mu^{6}M_{\rm BH}^{5}$, which motivates the adoption of an effective rate
\begin{equation}
\Gamma(\mu,M_{\rm BH})=16\,\mu^{6}\,M_{\rm geo}^{5},
\label{eq:Gamma}
\end{equation}
where
\begin{equation}
M_{\rm geo}=\frac{G\,M_{\rm BH}}{c^{3}}
\label{eq:Mgeo}
\end{equation}
where $M_{\rm geo}$ denotes the BH mass expressed in geometric time units. In conventional units, $GM_{\rm BH}/c^3$ has dimensions of time, while in geometrized units $G=c=1$ mass, length, and time are represented by the same quantity. This choice is convenient here because the boson mass $\mu$ is converted to inverse-time units, so that the combination $\mu M_{\rm geo}$ is dimensionless and $\Gamma$ has dimensions of an inverse time. Although the parameter scan is naturally
labeled by $\mu$ in eV, the numerical implementation consistently
converts this quantity to inverse-time units, so that $\Gamma$ has
dimensions of a rate throughout. Equation~(\ref{eq:Gamma}) should
be understood as an effective phenomenological prescription that
captures the strong parametric dependence on the dimensionless
coupling $\mu M_{\rm BH}$, rather than as a first-principles
description of the coupled Einstein-Klein-Gordon dynamics of the
cloud, which would in general also depend on the internal structure,
self-gravity, and formation history of the scalar configuration.

The scalar mass transfer channel is defined as
\begin{equation}
\dot M_{\rm scalar}
=\min\!\left[
\Gamma(\mu,M_{\rm BH})\,M_{c},\,
\Gamma_{\rm cap}\,M_{c}
\right],
\label{eq:mscalar}
\end{equation}
whenever $M_{c}>0$, and $\dot M_{\rm scalar}=0$ once the cloud reservoir is essentially exhausted. The parameter $\Gamma_{\rm cap}$ is a phenomenological upper bound on the scalar depletion rate. In all production runs we set $\Gamma_{\rm cap}=10^{-12}\mathrm{s}^{-1}$. Its role is to regularize numerically extreme regimes in which the effective transfer rate $\Gamma M_{c}$ would otherwise become excessively large over very short time intervals, preventing the stiff ODE integrator from being driven into regions where the effective description breaks down.  This cap should not be interpreted as a fundamental physical limit; rather, it is part of the effective prescription adopted here and marks the boundary at which a fully non-linear treatment would be required to follow the subsequent dynamics consistently. A dedicated $\Gamma_{\rm cap}$ variation study is left for future work. 

The mass balance equations for the coupled system then read
\begin{align}
\dot M_{\rm BH} &= \dot M_{\rm scalar}+\dot M_{\rm b,acc},
\label{eq:MBHdot}\\
\dot M_{c}     &= -\dot M_{\rm scalar},
\label{eq:Mcdot}\\
\dot M_{b}     &= -\dot M_{\rm b,acc},
\label{eq:Mbdot}
\end{align}
which ensure global mass conservation between the BH and the two matter reservoirs. Using the time-redshift relation Eq.~(\ref{eq:dtdz}), the corresponding evolution equations in redshift take the form
\begin{align}
\frac{dM_{\rm BH}}{dz}
&=-\frac{1}{(1+z)H(z)}
\left(\dot M_{\rm scalar}+\dot M_{\rm b,acc}\right),
\label{eq:dMBHdz}\\
\frac{dM_{c}}{dz}
&=\frac{1}{(1+z)H(z)}\,\dot M_{\rm scalar},
\label{eq:dMcdz}\\
\frac{dM_{b}}{dz}
&=\frac{1}{(1+z)H(z)}\,\dot M_{\rm b,acc}.
\label{eq:dMbdz}
\end{align} 
The positive signs appearing in Eqs.~(\ref{eq:dMcdz}) and (\ref{eq:dMbdz}) simply reflect the fact that the integration variable decreases with cosmic time: as the Universe evolves from high to low redshift, both reservoirs are consumed even though their derivatives with respect to $z$ are formally positive. The system~(\ref{eq:dMBHdz})-(\ref{eq:dMbdz}) constitutes the closed set of coupled nonlinear equations that is solved throughout the present study, once the initial conditions $M_{\rm BH}(z_{\rm ini})=M_{0}$, $M_{c}(z_{\rm ini})=M_{c,0}$, and $M_{b}(z_{\rm ini})=M_{b,0}$ have been specified at an initial high-redshift value $z_{\rm ini}$.

A number of physical observables can be extracted from the integrated trajectories. Two of them are particularly informative and will be systematically used in the analysis. The first is the BH mass evaluated at a fixed reference redshift, which provides a direct measure of how much of the surrounding matter has been accreted by a given cosmological epoch. The second is the arrival redshift $z_{\rm reach}(M_{\ast})$, defined as the redshift at which the BH mass first reaches a chosen threshold, and used in particular for $M_{\ast}=10^{9}\,M_\odot$ as a proxy for the onset of the SMBH regime. Complementarily, the total scalar and baryonic masses absorbed over the evolution, together with their ratio $\Delta M_{\rm scalar}/\Delta M_{\rm baryon}$, provide a clean diagnostic of which channel dominates the growth history. The Eddington luminosity along the trajectory,
\begin{equation}
L_{\rm Edd}(M_{\rm BH})
=\frac{4\pi G M_{\rm BH} m_{p} c}{\sigma_{T}},
\label{eq:LEdd}
\end{equation}
where $m_{p}$ is the proton mass and $\sigma_{T}$ the Thomson cross section, is also computed self-consistently and can be directly compared with the observed high-redshift quasar luminosity function.

Finally, the framework rests on a few assumptions that are worth
making explicit. The scalar cloud is assumed to exist as a
pre-formed effective reservoir at the initial redshift, without
attempting to model its cosmological formation or its internal
radial profile; back-reaction of the scalar transfer on the
surrounding geometry is not explicitly resolved; the BH is
treated as non-rotating, so that superradiant amplification is
absent and the scalar channel operates purely through
quasibound-state decay; and the baryonic reservoir is regarded
as a lumped mass that feeds the BH through an Eddington-like
prescription, without following the detailed hydrodynamics of
its inflow. These simplifications are what allow a broad,
tractable exploration of the four-dimensional parameter space
$(\mu,M_{0},M_{c,0},M_{b,0})$; a more complete relativistic and
astrophysical treatment is left for future work.

% ---
\section{Numerical implementation}
\label{sec:numerics}
% ---
The coupled system of ordinary differential equations~(\ref{eq:dMBHdz})-(\ref{eq:dMbdz}) is nonlinear and, in several regions of the explored parameter space, markedly stiff. The stiffness originates from the strong parametric dependence of the effective scalar rate $\Gamma\propto\mu^{6}M_{\rm BH}^{5}$ on both the boson mass and the growing BH mass: in the most efficient absorption regimes, the characteristic timescale of the scalar channel can become many orders of magnitude shorter than the Hubble time, while remaining fully coupled to the slower baryonic channel. A numerical scheme suited to stiff systems is therefore required in order to track accurately both the rapid scalar-depletion phases and the subsequent slow, Eddington-driven growth within the same integration.

The evolution is performed directly in redshift, adopting $z$ as the independent variable in the integrator. This choice presents two practical advantages. First, it removes the need to reconstruct redshift a posteriori from cosmic time, avoiding additional interpolation errors when comparing the numerical trajectories with observational quantities naturally expressed in terms of $z$. Second, it allows us to impose the final boundary at $z_{\rm end}=0$ directly and to identify diagnostics such as the arrival redshift $z_{\rm reach}(M_{\ast})$ along the solution itself, without any post-processing conversion.

In all scans presented in this work, the integration runs from $z_{\rm ini}=120$ to $z_{\rm end}=0$. This initial redshift is chosen to lie sufficiently early for all configurations considered, ensuring that the full relevant scalar and baryonic evolution is captured within the integration domain.

The system is integrated with the Python routine \texttt{solve\_ivp} from the \texttt{scipy.integrate} library, using the implicit \emph{Radau} method, a fifth-order implicit Runge--Kutta scheme of the Radau IIA family designed for stiff initial-value problems. For each individual run, the state vector $Y(z)$ is evaluated on a fixed grid of $5000$ points uniformly distributed between $z_{\rm ini}$ and $z_{\rm end}$, ensuring homogeneous coverage of the full evolution and facilitating the subsequent construction of interpolated observables. The relative and absolute integration tolerances are set to ${\tt rtol}=10^{-10}$ and ${\tt atol}=10^{-12}$, respectively, so the numerical error remains subdominant to the uncertainties inherent in the effective phenomenological description. Along each trajectory, we evaluate the Eddington luminosity in Eq.~(\ref{eq:LEdd}) on the same redshift grid and store it with the state vector. Regarding the possible formation of scalar clouds at high redshifts, we have added an appendix addressing the topic, Appendix~\ref{app:clouds}. However, a priori, no bound appears to restrict the possibility that scalar cloud formation processes could occur at $z \le 120$ within our study window for $\mu$.

An important element of the numerical setup is the treatment of
the depletion of the two reservoirs. In principle, the finite
mass constraints in Eqs.~(\ref{eq:mbacc}) and (\ref{eq:mscalar})
would make the right-hand sides of the evolution equations
non-smooth at the exact points where either $M_{c}$ or $M_{b}$
vanishes, potentially degrading the performance of the implicit
integrator. To avoid this, the integration is terminated as
soon as both reservoirs are found to be numerically negligible.
The termination criterion is
\begin{equation}
M_{c}\le 10^{-12}\,M_{c,0}\quad\text{and}\quad
M_{b}\le 10^{-12}\,M_{b,0},
\label{eq:depletion}
\end{equation}
that is, both reservoirs are required to have fallen by twelve
orders of magnitude below their initial values before the run
is stopped. At that point, the solution arrays are truncated,
so that no computational effort is wasted on integrating a
system that is dynamically inactive from that redshift onward.
This threshold is small enough that its precise value does not
affect any of the physical diagnostics considered here, but it
is nevertheless part of the numerical setup and is documented
explicitly in the interest of reproducibility.

A parameter scan is then performed over the four-dimensional
initial-condition space $(\mu,M_{0},M_{c,0},M_{b,0})$. The
boson mass is varied over a broad range covering the ultralight
sector,
$\mu\in[10^{-22},\,10^{-16}]\,{\rm eV}$, discretized
logarithmically to sample uniformly the exponent of the
Compton wavelength. The initial BH seed mass is likewise
sampled logarithmically across the physically motivated interval
$M_{0}\in\{10^{2},\,10^{3},\,10^{4},\,10^{5}\}\,M_\odot$,
which spans the transition between light Population~III-like
seeds and heavy direct-collapse-like seeds. The initial scalar
cloud mass is scanned over
$M_{c,0}\in\{10^{7},\,10^{8},\,10^{9},\,10^{10}\}\,M_\odot$,
covering a range that is comparable with typical dark-matter
halo masses at high redshift, while the initial baryonic
reservoir is treated in the same way. Every combination of
these parameters defines an independent integration; the full
set of runs then provides a discrete but comprehensive sampling
of the model's behavior across parameter space. The number of
runs, several thousand in total, is chosen to be large enough
to ensure that the trends and structures identified in the
diagnostic plots are not sensitive to the sampling density.

For each individual run, a set of derived quantities is
computed and stored. These include: (i) the three components of
$Y(z)$ on the reference grid, from which any redshift-resolved
observable can be reconstructed by interpolation; (ii) the
arrival redshift $z_{\rm reach}(M_{\ast})$ for the reference
threshold $M_{\ast}=10^{9}\,M_\odot$, obtained by locating the
zero of $M_{\rm BH}(z)-M_{\ast}$ along the trajectory; (iii)
the total masses transferred through each channel,
$\Delta M_{\rm scalar}$ and $\Delta M_{\rm baryon}$, together
with their ratio; and (iv) the redshift-dependent Eddington
luminosity $L_{\rm Edd}(z)$. These quantities are the
building blocks used in the subsequent statistical analysis
and in the comparison with the observational quasar sample.

The output of the scan is finally structured as a single
tabular dataset in which each row corresponds to one numerical
run and each column to one of the derived quantities above,
together with the initial conditions and the fiducial
cosmological parameters. This organization ensures a fully
reproducible analysis pipeline and makes it possible to
perform the various multidimensional cuts and projections that
underlie the results discussed in the following sections. All
plots and statistical diagnostics presented below are derived
from this common dataset, without any additional
post-processing beyond standard interpolation, binning, and
ranking operations.

\section{Results}\label{sec:results}

% --
\subsection{Illustrative redshift evolutions}
% --

\begin{figure*}[t]
    \centering

    \includegraphics[width=0.49\linewidth]{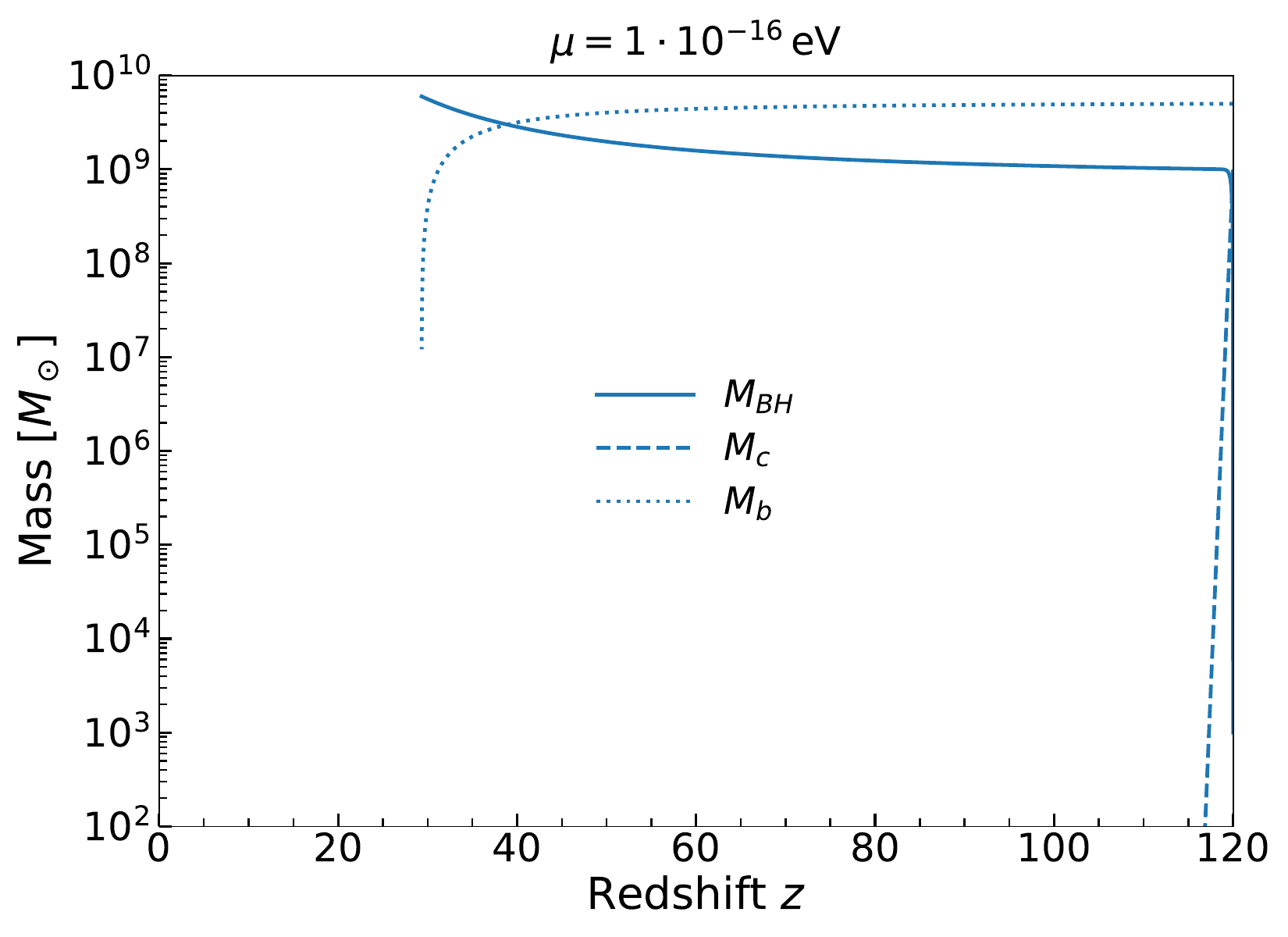}
    \hfill
    \includegraphics[width=0.49\linewidth]{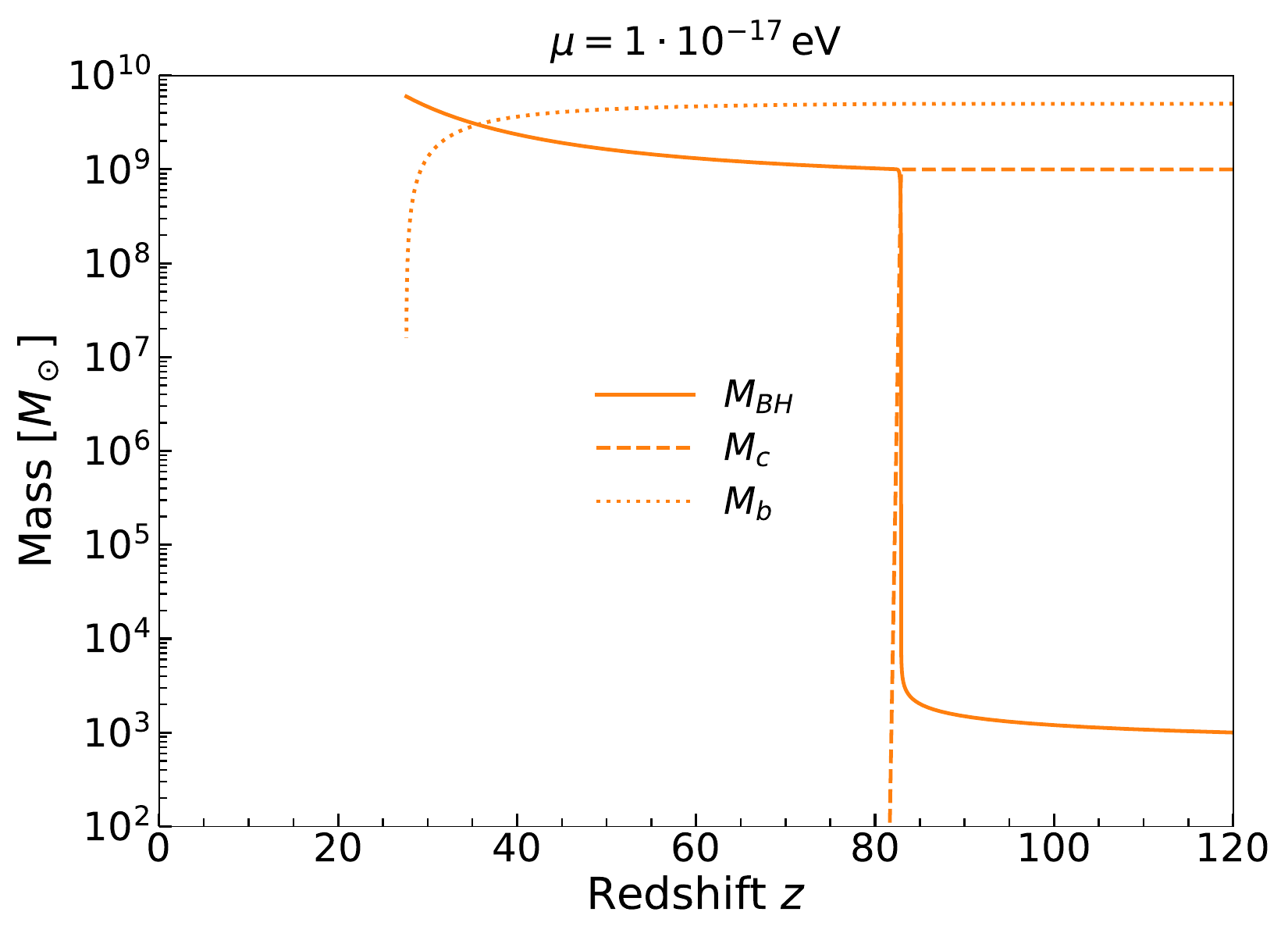}

    \vspace{0.1cm}

    \includegraphics[width=0.49\linewidth]{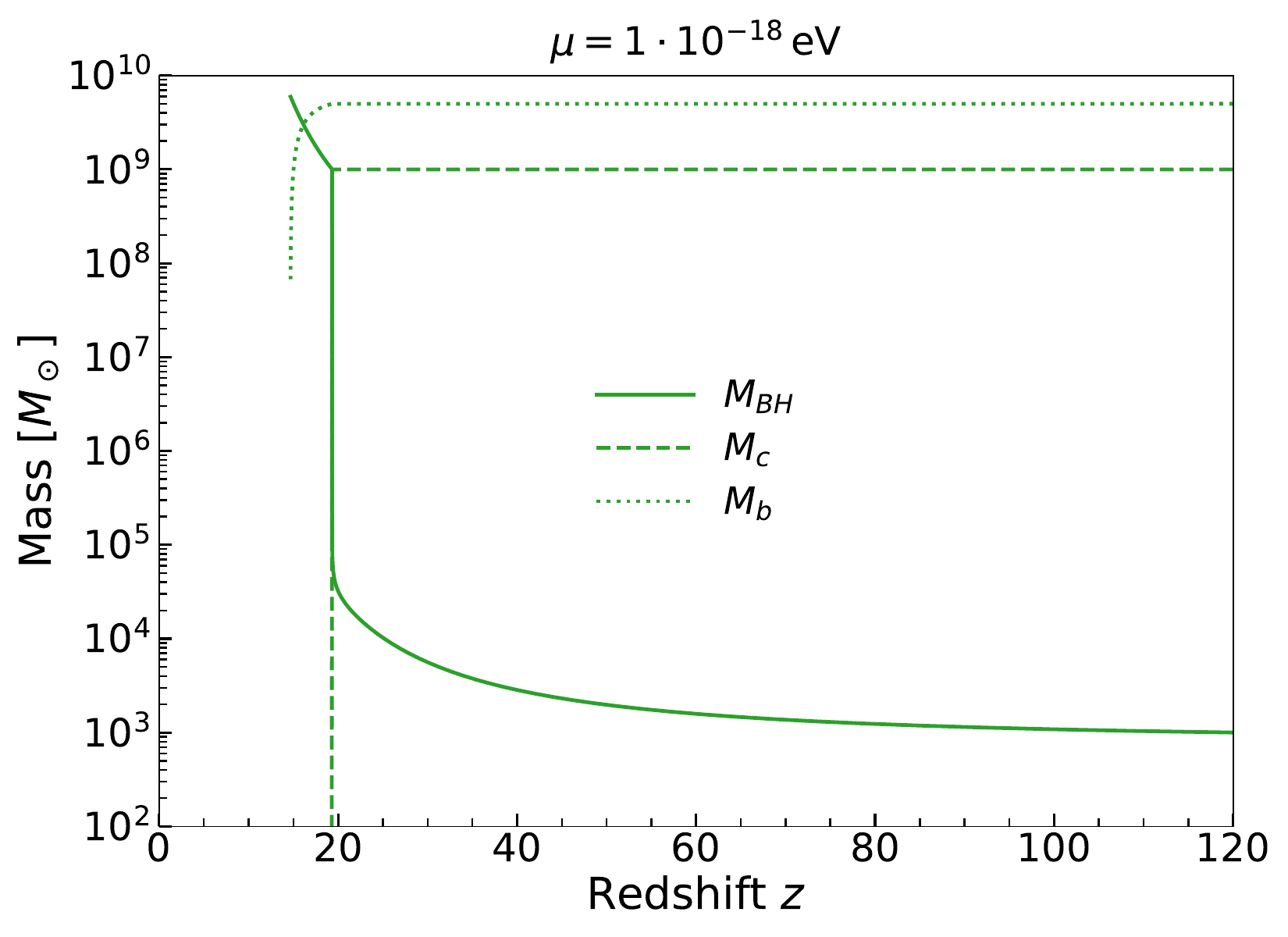}
    \hfill
    \includegraphics[width=0.49\linewidth]{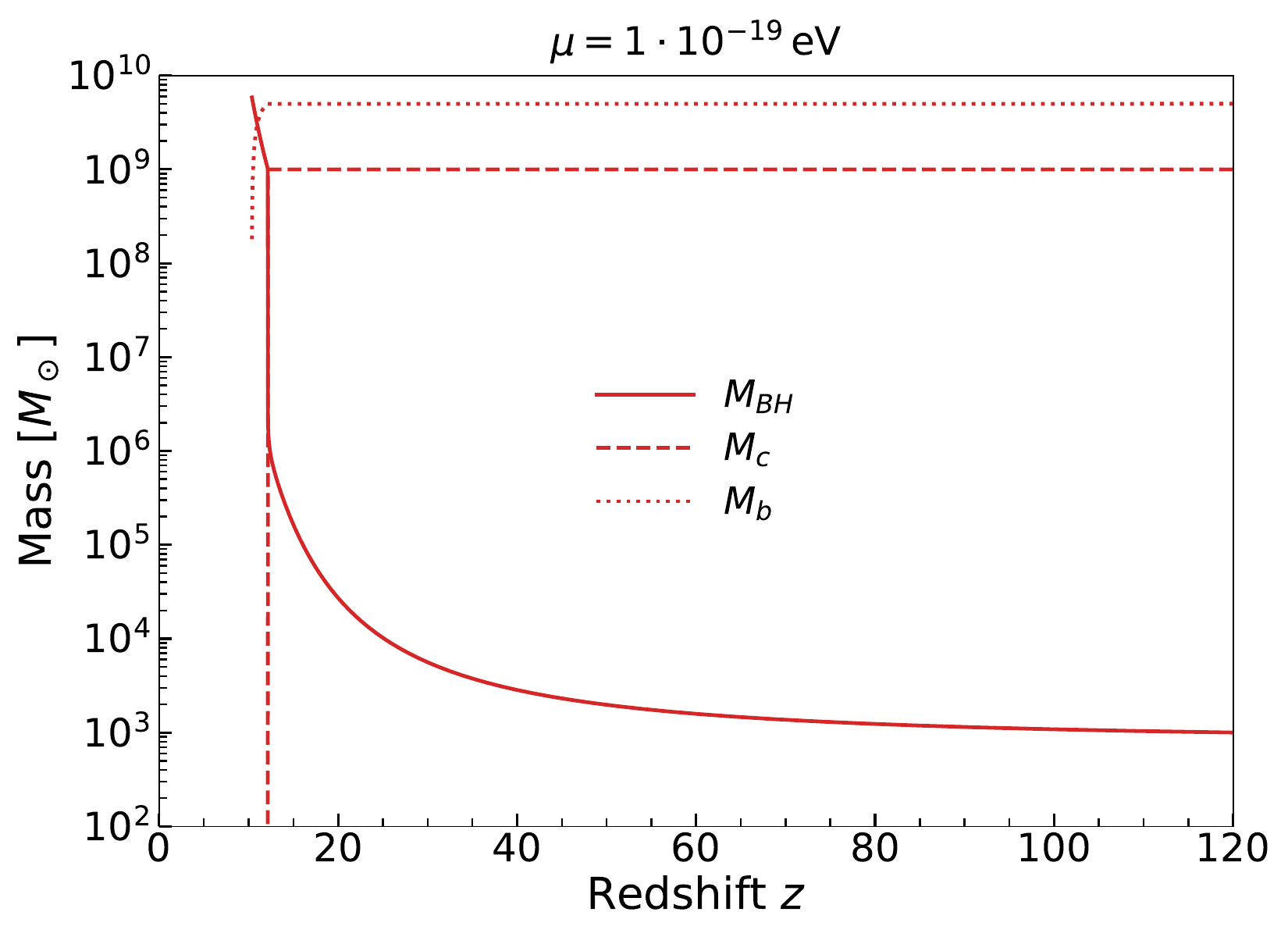}

    \caption{Representative redshift evolution of $M_{\rm BH}$ (solid), $M_c$ (dashed), and $M_b$ (dotted) for $M_{c,0}=10^9\,M_\odot$, with fixed $M_0=10^3\,M_\odot$ and $M_{b,0}=5\times10^9\,M_\odot$, for four values of the boson mass $\mu$. Although the evolution is formally defined down to $z=0$, each trajectory is terminated once both matter reservoirs are depleted, since no further mass transfer occurs within the model. Extending the evolution beyond this point would therefore require introducing additional matter supply beyond the finite-reservoir prescription adopted here.}
    \label{fig1}
\end{figure*}

Before the global parameter scan, it helps to examine a small set of representative individual evolutions, as they provide the most direct picture of the underlying dynamics. Figure~\ref{fig1} shows the redshift evolution of the three mass components,
\begin{equation}
    M_{\rm BH}(z),\qquad M_c(z),\qquad M_b(z),
\end{equation}

for a fixed initial BH seed mass, $M_0=10^3\,M_\odot$, a fixed baryonic reservoir, $M_{b,0}=5\times10^9\,M_\odot$, and an initial scalar-cloud mass $M_{c,0}=10^9\,M_\odot$. The four panels correspond to four representative values of the boson mass $\mu$, so that the role of this parameter can be isolated while keeping the remaining initial conditions unchanged.

In each panel of Figure~\ref{fig1}, the solid curve represents the BH mass $M_{\rm BH}$, the dashed curve the remaining scalar-cloud mass $M_c$, and the dotted curve the remaining baryonic reservoir $M_b$. This representation allows us to simultaneously follow the transfer of mass between the two reservoirs and the growing BH. A rapid decrease of $M_c$ signals an efficient transfer of scalar matter into the BH, whereas the smoother decrease of $M_b$ reflects the gradual consumption of the baryonic reservoir through the Eddington-like accretion channel. The evolution of $M_{\rm BH}$ therefore records the combined contribution of both feeding mechanisms.

Since the horizontal axis corresponds to redshift, larger values of $z$ equate to earlier cosmic times. Thus, the physical evolution proceeds from right to left toward the present epoch. Comparing the four panels immediately shows that the boson mass strongly controls both the onset and the efficiency of the scalar-assisted growth phase. For some values of $\mu$, the scalar cloud undergoes a rapid depletion episode at very high redshift, transferring a substantial fraction of its mass to the BH over a comparatively short cosmological interval. Once the cloud has been depleted, the subsequent evolution becomes predominantly baryonic. For other values of $\mu$, scalar absorption is delayed or considerably less efficient, and the BH remains closer to the baryonic-dominated growth regime for a much larger fraction of its evolution.

Figure~\ref{fig1} therefore provides a direct illustration of the nonlinear mechanism underlying the global trends discussed below. The boson mass determines how efficiently and at what epoch the scalar channel becomes active, while the finite scalar and baryonic reservoirs determine how much mass can ultimately be supplied through each channel and for how long the corresponding growth phase can be sustained. In particular, the rapid depletion episodes visible in some of the panels illustrate how a configuration that initially evolves only slowly can enter a regime of highly efficient scalar absorption once the appropriate conditions are reached.
In Table \ref{tab:depletion_redshifts_both} we show the depletion redshifts of both clouds for each value of $\mu$ corresponding to the point where all the mass has been accreted by the BHs.

\begin{table}[t]
\centering
\caption{Depletion and stopping redshifts for the Fig.~\ref{fig1} runs
($M_0=10^3\,M_\odot$, $M_{c,0}=10^9\,M_\odot$, $M_{b,0}=5\times10^9\,M_\odot$).
Here, $z_{\rm dep}^{(c)}$ is defined by $M_c \le 10^{-12}M_{c,0}$ and
$z_{\rm dep}^{(b)}$ by $M_b \le 10^{-12}M_{b,0}$, while $z_{\rm stop}$ is the
last redshift reached by the numerical integrator.}
\label{tab:depletion_redshifts_both}
\begin{tabular}{cccc}
\hline
$\mu$ [eV] & $z_{\rm dep}^{(c)}$ & $z_{\rm dep}^{(b)}$ & $z_{\rm stop}$ \\
\hline
$10^{-16}$ & 114.751 & 29.314 & 29.314 \\
$10^{-17}$ & 80.764  & 27.607 & 27.607 \\
$10^{-18}$ & 19.270  & 14.670 & 14.670 \\
$10^{-19}$ & 12.135  & 10.346 & 10.346 \\
\hline
\end{tabular}
\end{table}

It is important to clarify a numerical feature associated with the high-$\mu$ regime. As seen in the previous figures, increasing the boson mass shifts the onset of efficient scalar absorption toward progressively higher redshifts. For $\mu=10^{-16}\,\mathrm{eV}$, the transfer is effectively immediate within the integration domain, occurring already at the initial redshift $z_{\rm ini}$. We have verified that this behavior persists when the integration is started at even higher values of $z_{\rm ini}$, indicating that, for such large $\mu$, the apparent absorption redshift is set by the chosen starting epoch and should therefore be interpreted as a numerical upper-limit effect rather than as a physically resolved transition. As $\mu$ is decreased, however, the onset of the rapid-absorption phase moves away from the numerical boundary and develops a genuine dependence on the boson mass. For example, for $\mu=10^{-17}\mathrm{eV}$, the scalar cloud is rapidly absorbed at very high but finite redshift, around $z\sim80$. Further decreasing $\mu$ shifts this transition systematically toward lower redshifts. Thus, while the largest boson masses considered here saturate the integration's temporal resolution at early times, the lower-$\mu$ cases clearly reveal the expected trend: smaller boson masses delay the onset of the efficient scalar-transfer regime. 

For clarity, in the main text we show the representative case $M_{c,0}=10^9M_\odot$, while analogous evolutions for the remaining values of $M_{c,0}$ considered in the parameter scan are presented in Appendix~\ref{app:trajplots}. As will become clearer from the global analysis, the rapid scalar-absorption regimes illustrated here are especially relevant for the high-redshift evolution of the most efficiently growing configurations, where the scalar channel can induce a substantial increase in the BH mass before ordinary baryonic accretion becomes dominant.

% --
\subsection{Early-SMBH formation and arrival redshift}
% --

\begin{figure}[t]
    \centering
    \includegraphics[width=\columnwidth]{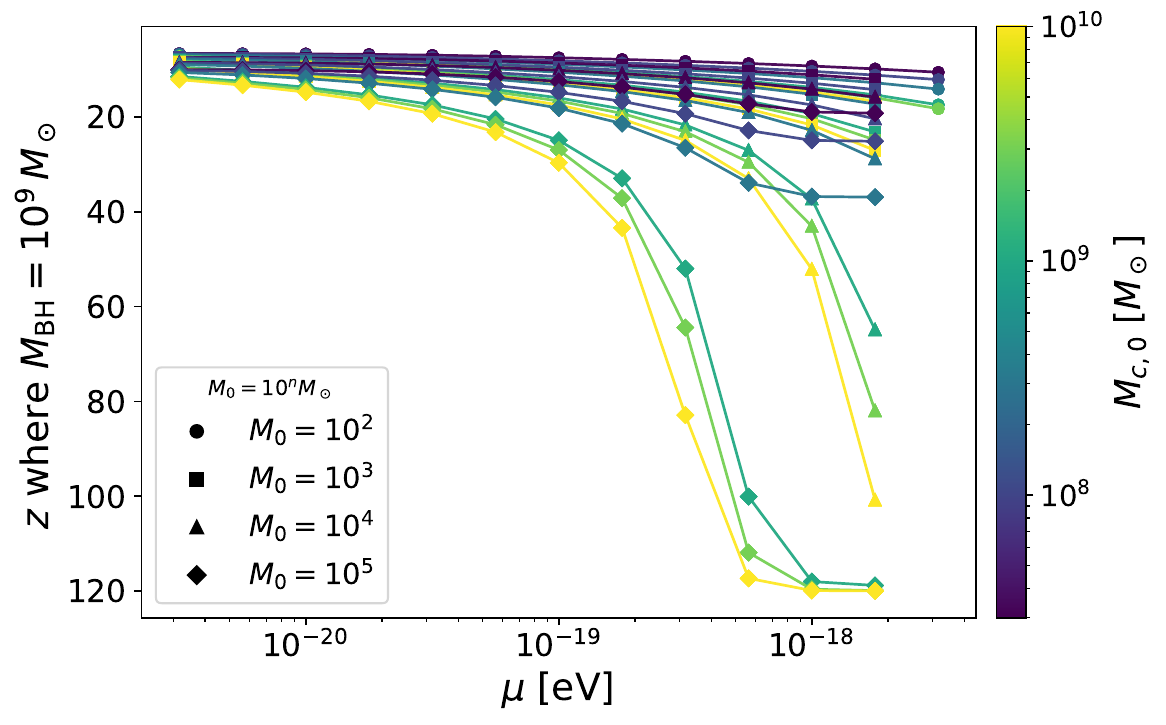}
    \caption{Arrival redshift $z_{\rm reach}(10^9\,M_\odot)$ as a
    function of the boson mass $\mu$. Different markers distinguish
    the initial BH seed mass $M_0$, while the color scale encodes the
    initial scalar-cloud mass $M_{c,0}$.}
    \label{fig:zreach_main}
\end{figure}

We now present a systematic parametric study of how the initial
conditions affect the subsequent evolution of the system. In
Fig.~\ref{fig:zreach_main}, we show the redshift at which the BH first
reaches a mass of $M_{\rm BH}=10^9\,M_\odot$. The results are displayed
as a function of the boson mass $\mu$, while different markers
distinguish the initial BH seed mass $M_0$ and the color scale encodes
the initial scalar-cloud mass $M_{c,0}$. Although the figure combines
several dimensions of the parameter space, a number of clear trends
can be identified.
The first noticeable feature is the larger concentration of solutions at relatively low redshifts. According to the convention adopted in the figure, lower values of $z$ appear toward the upper part of the vertical axis. Therefore, these configurations correspond to systems that require a longer cosmological time to reach $M_{\rm BH}=10^9\,M_\odot$. This indicates that, over a substantial fraction of the explored parameter space, the combined growth process is comparatively slow. A similar trend can be identified from the color coding. The darker points, corresponding to smaller initial scalar-cloud masses, preferentially cluster in this low-$z$ region and therefore describe comparatively inefficient scalar-assisted growth. As the color becomes progressively lighter, corresponding to larger values of $M_{c,0}$, the solutions extend toward substantially higher redshifts, reaching values as high as $z\sim120$ for the most extreme initial conditions considered, which are low scalar particle mass $\mu$, large BH seed and scalar cloud masses.

The dependence on $M_{c,0}$ can be understood naturally. A more massive initial cloud provides a larger reservoir of scalar-field energy that can potentially be transferred to the BH. Consequently, configurations with larger $M_{c,0}$ are generally able to reach the target mass at earlier cosmological times, i.e., at larger redshifts. Nevertheless, Fig.~\ref{fig:zreach_main} also contains configurations with large $M_{c,0}$ that reach the target mass only at comparatively low redshift. The availability of a massive scalar reservoir is therefore a necessary ingredient for a large scalar contribution, but is not by itself sufficient to guarantee an efficient early-growth phase. 

The second parameter, namely the boson mass $\mu$, plays, as seen before, a crucial role. The configurations reaching $M_{\rm BH}=10^9\,M_\odot$ at the highest redshifts (earliest cosmological times) are confined to a comparatively narrow region of the parameter space explored here. In particular, for $\mu\lesssim10^{-19}\,{\rm eV}$, scalar absorption is strongly suppressed and the target mass is generally reached only at lower redshifts. As $\mu$ increases within the range considered here, the scalar-transfer rate becomes more efficient and, provided that a sufficiently massive cloud is available, the BH can reach $10^9\,M_\odot$ substantially earlier. The boson mass and the initial cloud mass therefore play complementary roles: $\mu$ controls the characteristic transfer rate, whereas $M_{c,0}$ determines the amount of scalar matter available once the efficient-transfer regime is reached.

The strong dependence on $\mu$ can be understood in terms of the characteristic length scales of the scalar field and the BH. The reduced Compton wavelength associated with a boson of mass $\mu$ is
\begin{equation}
    \bar{\lambda}_{\rm C}
    =
    \frac{\hbar}{\mu c},
    \label{eq:compton_wavelength}
\end{equation}
while the gravitational length scale of a BH of mass $M_{\rm BH}$ is
\begin{equation}
    r_g
    =
    \frac{G M_{\rm BH}}{c^2}.
    \label{eq:gravitational_radius}
\end{equation}
The ratio between these two scales defines the dimensionless gravitational coupling
\begin{equation}
    \alpha
    \equiv
    \frac{r_g}{\bar{\lambda}_{\rm C}}
    =
    \frac{G M_{\rm BH}\mu}{\hbar c},
    \label{eq:gravitational_coupling}
\end{equation} 
which reduces to $\alpha=\mu M_{\rm BH}$ in geometrized units. Thus, the boson mass does not determine the efficiency of the process independently of the BH mass. Rather, both quantities jointly determine the hierarchy between the scalar wavelength and the BH scale. In the small-coupling regime, $\alpha\ll1$, the scalar wavelength is much larger than the characteristic BH scale and absorption by the horizon is strongly suppressed. As the BH grows, $\alpha$ increases, the hierarchy between the two scales is reduced, and the scalar-transfer rate rises rapidly.

The scaling encoded in Eq.~(\ref{eq:gravitational_coupling}) can be made more transparent by expressing the boson and BH masses in terms of their characteristic orders of magnitude,
\begin{equation}
    \mu = 10^{-n}\,{\rm eV},
    \qquad
    M_{\rm BH}=10^m M_\odot.
\end{equation}
Substituting these expressions into the definition of the
gravitational coupling gives
\begin{equation}
    \alpha
    =
    \frac{G M_{\rm BH}\mu}{\hbar c}
    \simeq
    7.48\times10^{m-n+9}.
    \label{eq:alpha_mn}
\end{equation} 
Thus, the efficiency of the scalar-transfer channel is controlled not by $\mu$ or $M_{\rm BH}$ separately, but by the combination of their orders of magnitude through $m-n$. For a characteristic coupling $\alpha_\ast$ at which scalar absorption becomes dynamically relevant, the corresponding BH and boson mass scales satisfy approximately
\begin{equation}
    m
    \simeq
    n-9.87+\log_{10}\alpha_\ast.
    \label{eq:mn_scaling}
\end{equation}

This relation provides a simple interpretation of the boson-mass window identified in the numerical scan. For a fixed characteristic value of $\alpha_\ast$, increasing the BH mass by one order of magnitude shifts the corresponding efficient boson mass downward by one order of magnitude. Conversely, for a fixed BH mass, increasing $\mu$ moves the system toward larger coupling and therefore toward a more efficient scalar-transfer regime. In other words, the efficient window should not be understood as a universal interval in $\mu$ alone, but rather as the projection, onto the boson-mass axis, of a preferred range in the dimensionless combination $\mu M_{\rm BH}$.

As an order-of-magnitude illustration, if the relevant coupling is assumed to be of order unity, the condition $\alpha\sim1$ implies approximately
\begin{equation}
    m\simeq n-10.
\end{equation}
Thus, a boson mass around $\mu\sim10^{-15}\,{\rm eV}$ corresponds to a BH mass scale of order $M_{\rm BH}\sim10^{5}\,M_\odot$, while $\mu\sim10^{-18}\,{\rm eV}$ corresponds instead to $M_{\rm BH}\sim10^{8}\,M_\odot$. More generally, within the small-coupling regime relevant to the effective prescription employed here, the same mapping applies with the appropriate value of $\alpha_\ast<1$. This makes explicit why the preferred boson-mass region shifts as the BH grows and why different seed masses can enter the efficient scalar-absorption regime at different stages of their cosmological evolution.

This relation also clarifies the role of the initial BH mass $M_0$. As shown in Fig.~\ref{fig:zreach_main}, the markers associated with smaller values of $M_0$ generally fail to reach the target mass at redshifts as high as those attained by more massive seeds. This is partly because smaller seeds must accumulate more mass before reaching $10^9\,M_\odot$, but this is not the only effect. For the effective scalar rate adopted in this work, we have
\begin{equation}
    \Gamma
    =
    16\mu^6M_{\rm BH}^{5}
    =
    \frac{16\alpha^6}{M_{\rm BH}},
    \label{eq:Gamma_alpha_results}
\end{equation} 
in geometrized units. Consequently, the seed mass also determines the initial efficiency of the scalar-transfer channel. At fixed $\mu$, the initial rates associated with two seeds satisfy
\begin{equation}
    \frac{\Gamma_0^{(2)}}{\Gamma_0^{(1)}}
    =
    \left(
        \frac{M_0^{(2)}}{M_0^{(1)}}
    \right)^5.
    \label{eq:seed_rate_ratio}
\end{equation}
Within the regime of validity of the effective prescription, increasing the seed mass by one order of magnitude therefore enhances the initial scalar-transfer rate by five orders of magnitude. More massive seeds are not only closer to the target SMBH mass, but also begin their evolution with a substantially more efficient scalar channel.

Overall, the arrival-redshift analysis confirms that the bosonic component can substantially alter the epoch at which the BH reaches $10^9\,M_\odot$. At the same time, the dependence on the sampled parameters is sufficiently structured that no single optimal configuration should be over-interpreted. The robust conclusion is that scalar-assisted growth can advance the onset of the SMBH regime, but the magnitude of this effect depends sensitively on the boson mass, the initial seed mass, and the amount of scalar matter available.

For the mass scales explored here, the configurations producing the earliest SMBH assembly are preferentially concentrated in the boson-mass range
\begin{equation}
    \mu\sim10^{-19}\text{--}10^{-16}\,{\rm eV}.
\end{equation}
As discussed above, this interval should not be interpreted as a universal boson-mass window, but as the realization, for the BH masses considered here, of the more general scale-matching condition controlled by $\alpha=\mu M_{\rm BH}$. The corresponding efficient range therefore shifts toward smaller $\mu$ as the characteristic BH mass increases.

The combined dependence on $\mu$ and $M_{\rm BH}$ also explains the rapid scalar-absorption episodes seen in the representative evolutions of Fig.~\ref{fig1}. A cloud may initially evolve very slowly because the seed begins in a strongly suppressed small-$\alpha$ regime. However, baryonic accretion and the gradual transfer of scalar matter increase $M_{\rm BH}$, and hence also $\alpha$. Since $\Gamma\propto\alpha^6/M_{\rm BH}$, this produces a strong positive
feedback: the increase in the BH mass enhances the scalar-transfer
rate, which supplies additional mass to the BH and further accelerates
the depletion of the cloud. For suitable values of $\mu$, $M_0$, and
$M_{c,0}$, the system can therefore transition from an initially slow
evolution to a short phase of rapid scalar absorption, allowing the
SMBH mass scale to be reached at a significantly earlier cosmological
epoch.

\begin{figure*}[t]
    \centering

    \includegraphics[width=0.49\linewidth]{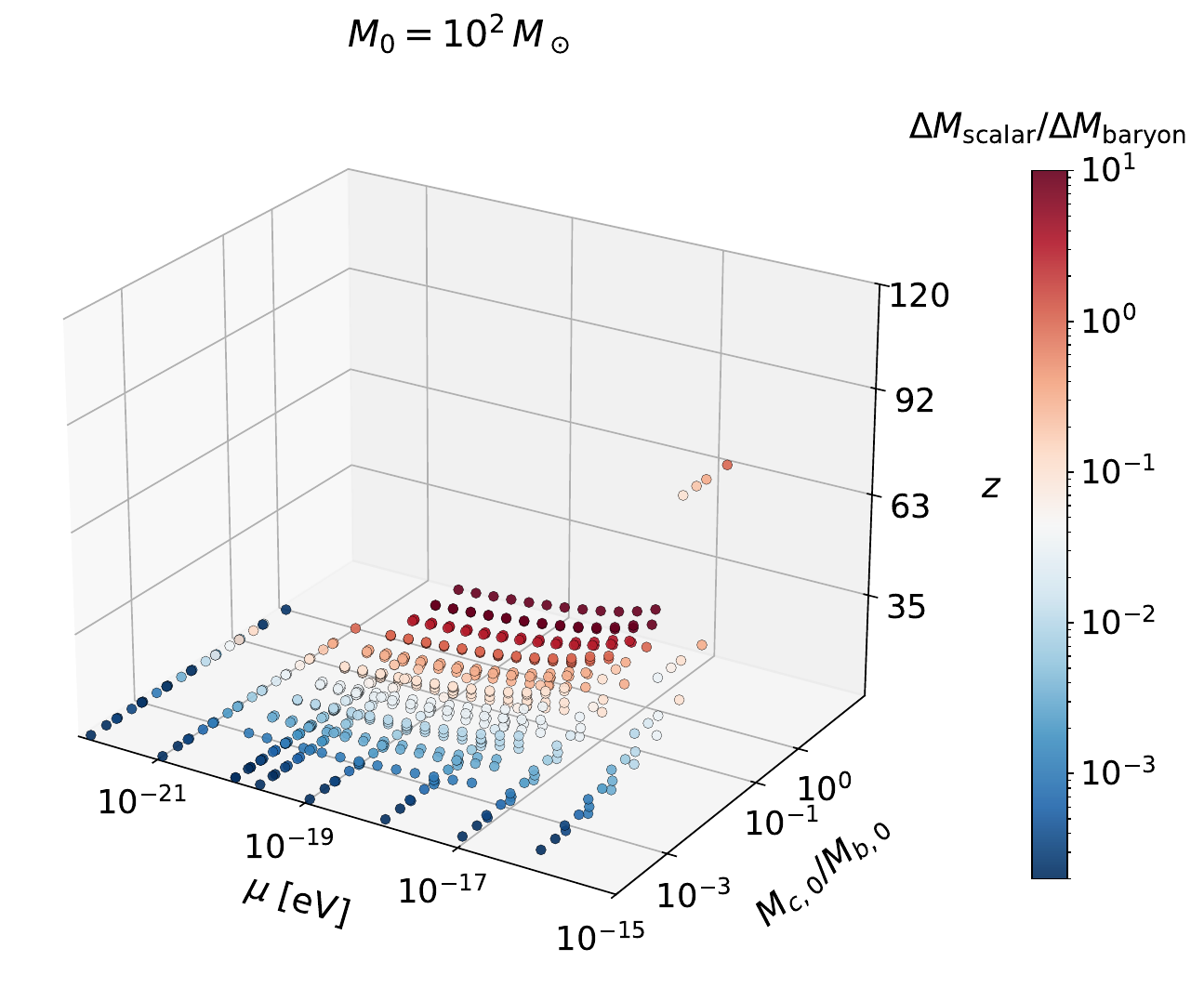}
    \hfill
    \includegraphics[width=0.49\linewidth]{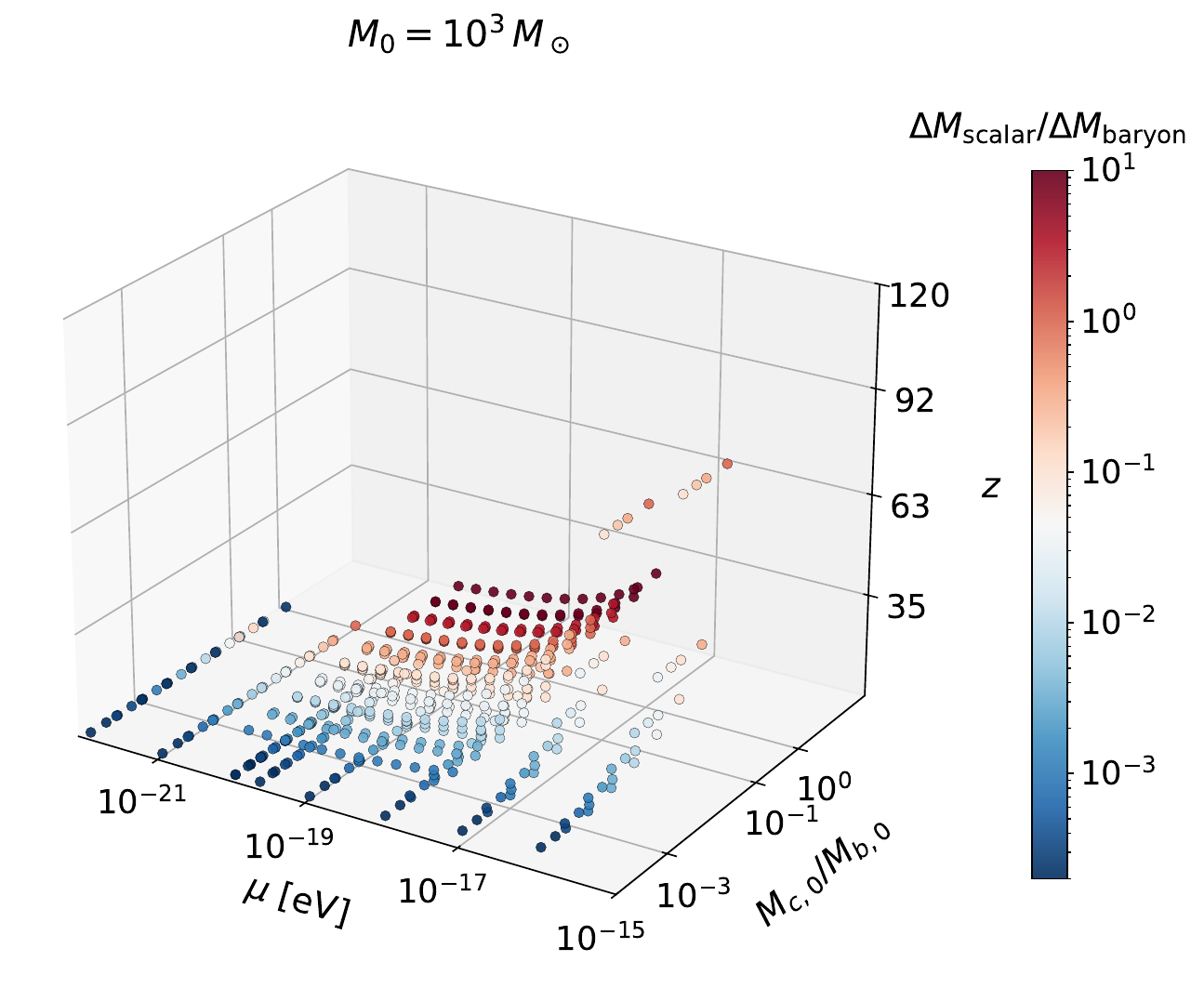}

    \vspace{0.1cm}

    \includegraphics[width=0.49\linewidth]{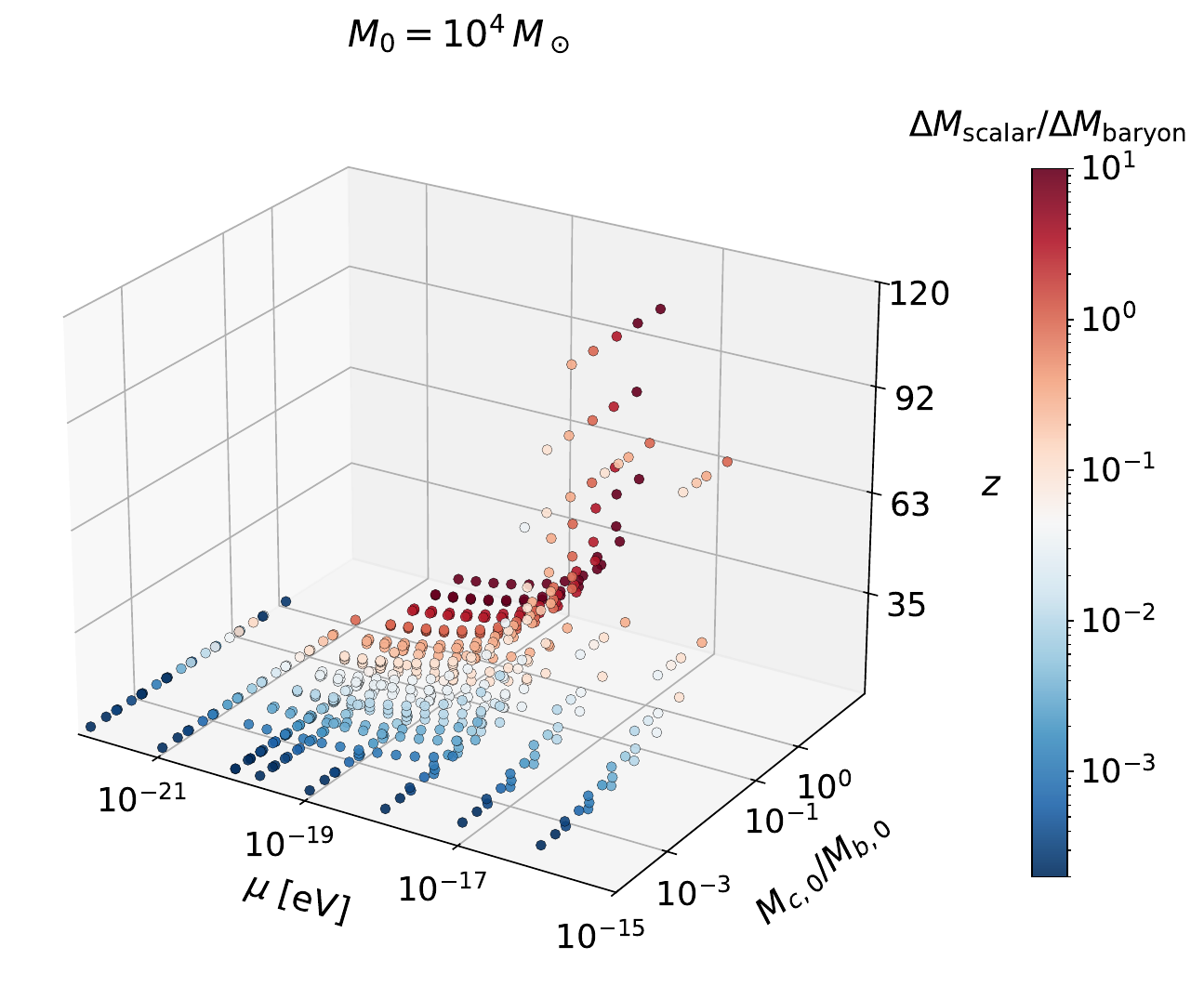}
    \hfill
    \includegraphics[width=0.49\linewidth]{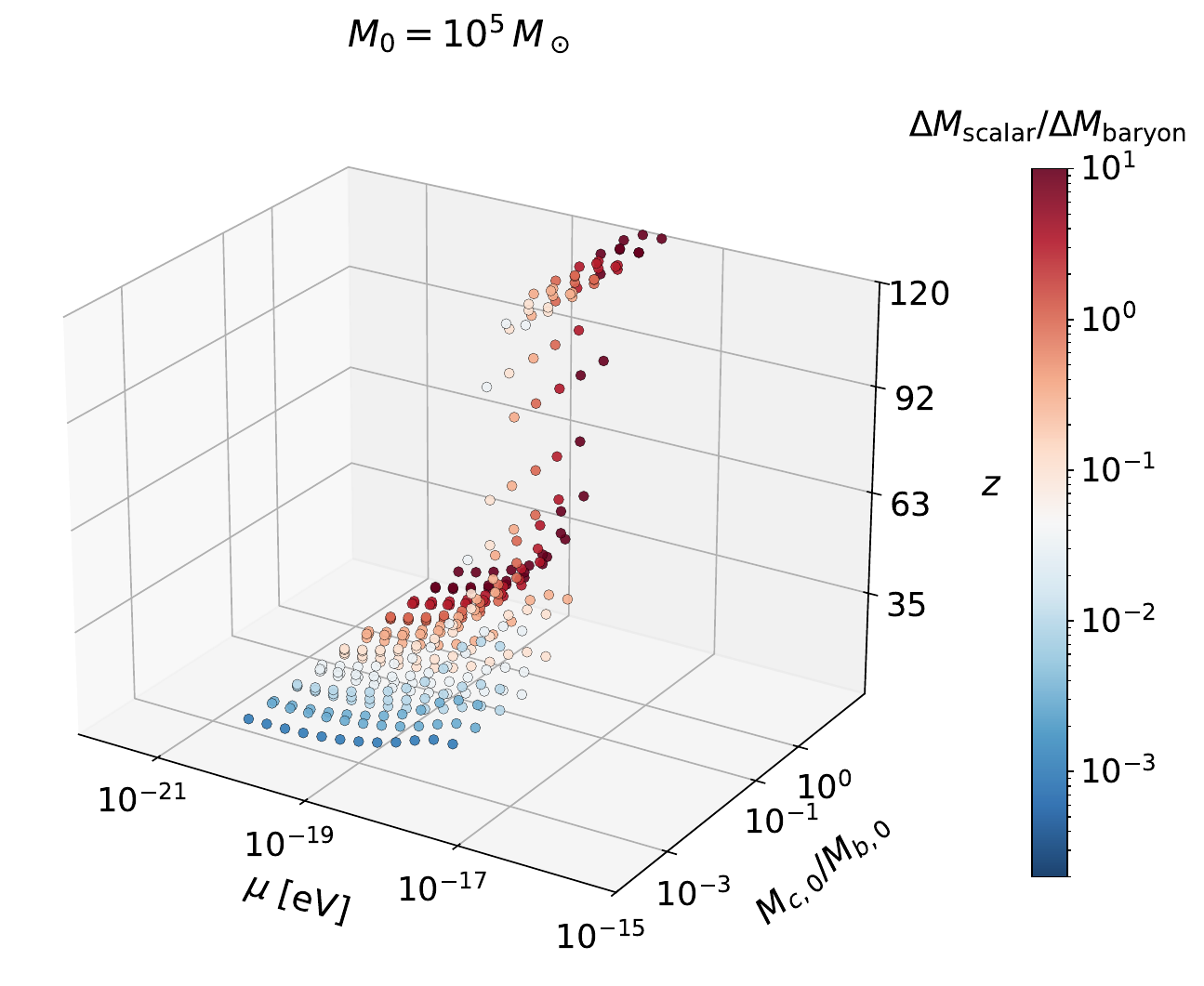}

    \caption{Multidimensional analysis for different starting $M_0$ BH seeds: upper left $M_0=100 M_{\odot}$, upper right $M_0=1000 M_{\odot}$, lower left $M_0=10^{4} M_{\odot}$ and lower right $M_0=10^{5} M_{\odot}$. The axes determine the mass of the boson $\mu[\mathrm{eV}]$, the ratio ${M_{c,0}}/{M_{b,0}}$ and the redshift $z$, while the colormap determines the amount of scalar over baryonic matter absorbed at the end stage of the system.  }
    \label{fig:four_panels}
\end{figure*}

Conversely, when the initial coupling
\begin{equation}
    \alpha_0
    =
    \frac{G M_0\mu}{\hbar c}
    \simeq
    7.5\times10^{-3}
    \left(\frac{M_0}{10^6\,M_\odot}\right)
    \left(\frac{\mu}{10^{-18}\,{\rm eV}}\right)
    \label{eq:alpha0_numerical}
\end{equation}
is too small, the scalar rate can remain suppressed over a substantial
fraction of the cosmological evolution. In this regime, baryonic
accretion remains the dominant growth channel, the scalar cloud is
incorporated only gradually, and the BH reaches the target mass at a
lower redshift. This expression also makes clear that comparable
initial couplings can be obtained from different combinations of
$\mu$ and $M_0$.

% --
\subsection{Full parametric analysis}
% --

Finally, we turn to a comprehensive multidimensional analysis in which all the relevant parameters of the system are considered simultaneously. This allows us to identify, in a much more direct way, the regions of parameter space associated with the highest growth efficiency and to determine the specific initial conditions under which these regimes arise. To this end, we analyze a dataset containing several thousand numerical solutions, providing a sufficiently broad sampling of the parameter space to extract robust trends.

Figure~\ref{fig:four_panels} shows four three-dimensional subplots, each corresponding to a different value of the initial BH mass $M_0$. Within each panel, every numerical solution is represented by a point in the three-dimensional space
\begin{equation}
P=P\left(\mu,\frac{M_{c,0}}{M_{b,0}},z\right),
\end{equation}
where $\mu$ is the boson mass, $M_{c,0}/M_{b,0}$ is the initial ratio between the scalar-cloud and baryonic reservoirs, and $z=z_{\rm reach}$ denotes the redshift at which the BH reaches $M_{\rm BH}=10^9\,M_\odot$. The color assigned to each point encodes the ratio
\begin{equation}
\frac{\Delta M_{\rm scalar}}{\Delta M_{\rm baryon}},
\end{equation}
which measures the relative contribution of the scalar and baryonic channels to the total mass accreted by the BH.

The quantity $M_{c,0}/M_{b,0}$ therefore characterizes the initial balance between the two available matter reservoirs, while $\Delta M_{\rm scalar}/\Delta M_{\rm baryon}$ quantifies their relative importance after the accretion process has taken place. This diagnostic naturally distinguishes three qualitatively different regimes:
\begin{equation}
\frac{\Delta M_{\rm scalar}}{\Delta M_{\rm baryon}}
\begin{cases}
<1,  \text{baryonic-accretion dominated},\\
\simeq 1,  \text{comparable contributions},\\
> 1,  \text{scalar-accretion dominated}.
\end{cases}
\label{eq:scalar_baryon_ratio}
\end{equation}
In this way, the figure simultaneously captures the dependence of the arrival redshift on the boson mass and on the initial matter composition, while also revealing which accretion channel dominates the final growth of the BH.
As discussed in the previous section, the BH reaches the reference mass at progressively higher redshifts as the initial seed mass $M_0$ increases, with both the boson mass $\mu$ and the amount of matter initially stored in the scalar reservoir playing a fundamental role. Figure~\ref{fig:four_panels} makes this behavior particularly clear by simultaneously showing the arrival redshift and the relative contributions of the scalar and baryonic accretion channels. In general, the configurations reaching $M_{\rm BH}=10^9\,M_\odot$ at the largest redshifts are associated with values of
\begin{equation}
\frac{\Delta M_{\rm scalar}}{\Delta M_{\rm baryon}}>1,
\end{equation}
and in many cases this ratio becomes significantly larger than unity. This indicates that the growth of the BH in the most efficient high-redshift scenarios is strongly dominated by the scalar channel, while the baryonic contribution becomes comparatively subdominant. These results reinforce the scale-dependent picture discussed in the previous subsection. In particular, the configurations exhibiting the largest scalar-assisted growth efficiencies are again concentrated, for the BH mass scales explored here, around $\mu\sim10^{-19}-10^{-16}\,{\rm eV}$. As emphasized above, this interval reflects the preferred range of the dimensionless coupling $\alpha=\mu M_{\rm BH}$ rather than a universal boson-mass window.

% ---------------------------------
\subsection{Comparison with observed high-redshift quasars}
% ---------------------------------

Beyond its intrinsic theoretical interest, the coupled scalar and baryonic growth model developed above must be confronted with the observed SMBH population in order to assess its cosmological relevance. This subsection is devoted to such a comparison, with the twofold aim of identifying regions of parameter space that are broadly compatible with the observed $(z,M_{\rm BH})$ distribution of quasars and of quantifying that compatibility through a well-defined statistical diagnostic. The comparison is performed by projecting the ensemble of numerical trajectories obtained from the parameter scan onto the observational plane, so that each configuration in the explored four-dimensional space $(\mu,M_0,M_{c,0},M_{b,0})$ is represented by its full redshift-resolved growth history and can be compared with the measured BH masses.

The parameter scan spans the four-dimensional space
\begin{equation}
\begin{split}
&\mu \in \left[10^{-21},\,10^{-16}\right]\,{\rm eV},
\\
&M_{0} \in \left[10^{2},\,10^{5}\right] M_{\odot},
\\
&M_{c,0} \in \left[3\times10^{7},\,10^{10}\right] M_{\odot},
\\
&M_{b,0} \in \left[3\times10^{8},\,3\times10^{10}\right] M_{\odot}.
\end{split}
\label{grid_vals}
\end{equation}
Each point in the resulting parameter grid is evolved
independently and subsequently compared with the observed
$(z,M_{\rm BH})$ distribution using the statistical metric defined
below.

The observational sample considered in this work consists of high-redshift active galactic nuclei and quasars with reliable virial or single-epoch BH mass estimates \cite{Sanchis-Gual:2021ufg,mazzucchelli2023xqr,Wang:2020zae,wang2018discovery,juodvzbalis2026direct,matsuoka2018subaru,mortlock2011luminous,yang2020poniua,Banados2018,2021ApJ...907L...1W,tripodi2024red,larson2023ceers,juodvzbalis2026jades}. After applying the selection and consistency filters adopted in the present analysis, the final catalog contains over $70000$ sources. The luminosity dataset used in Fig.~6 is drawn from this same merged catalog and corresponds to a restricted subsample with $4\leq z\leq 9$ and available luminosity-uncertainty estimates. The compilation spans approximately the redshift interval $z\sim1$-$9$, extending from the transition to the SMBH-dominated era down to the epoch at which the mass function of luminous quasars is best characterized. In the $(z,M_{\rm BH})$ plane, the sample forms a broad and densely populated cloud between $M_{\rm BH}\sim 10^6 M_\odot$ and $M_{\rm BH}\sim 10^{10} M_\odot$, with a marked concentration in the range $10^8$--$10^9 M_\odot$. Only a comparatively small number of extreme sources reach or exceed $10^9\,M_\odot$ at $z\gtrsim6$ \cite{mazzucchelli2023xqr,Wang:2020zae,wang2018discovery,juodvzbalis2026direct,matsuoka2018subaru,mortlock2011luminous,yang2020poniua,Banados2018,2021ApJ...907L...1W,tripodi2024red,larson2023ceers,juodvzbalis2026jades}, precisely the regime that provides the most stringent constraint on any theoretical model of early SMBH assembly.

\begin{figure}[t]
\includegraphics[width=\columnwidth]{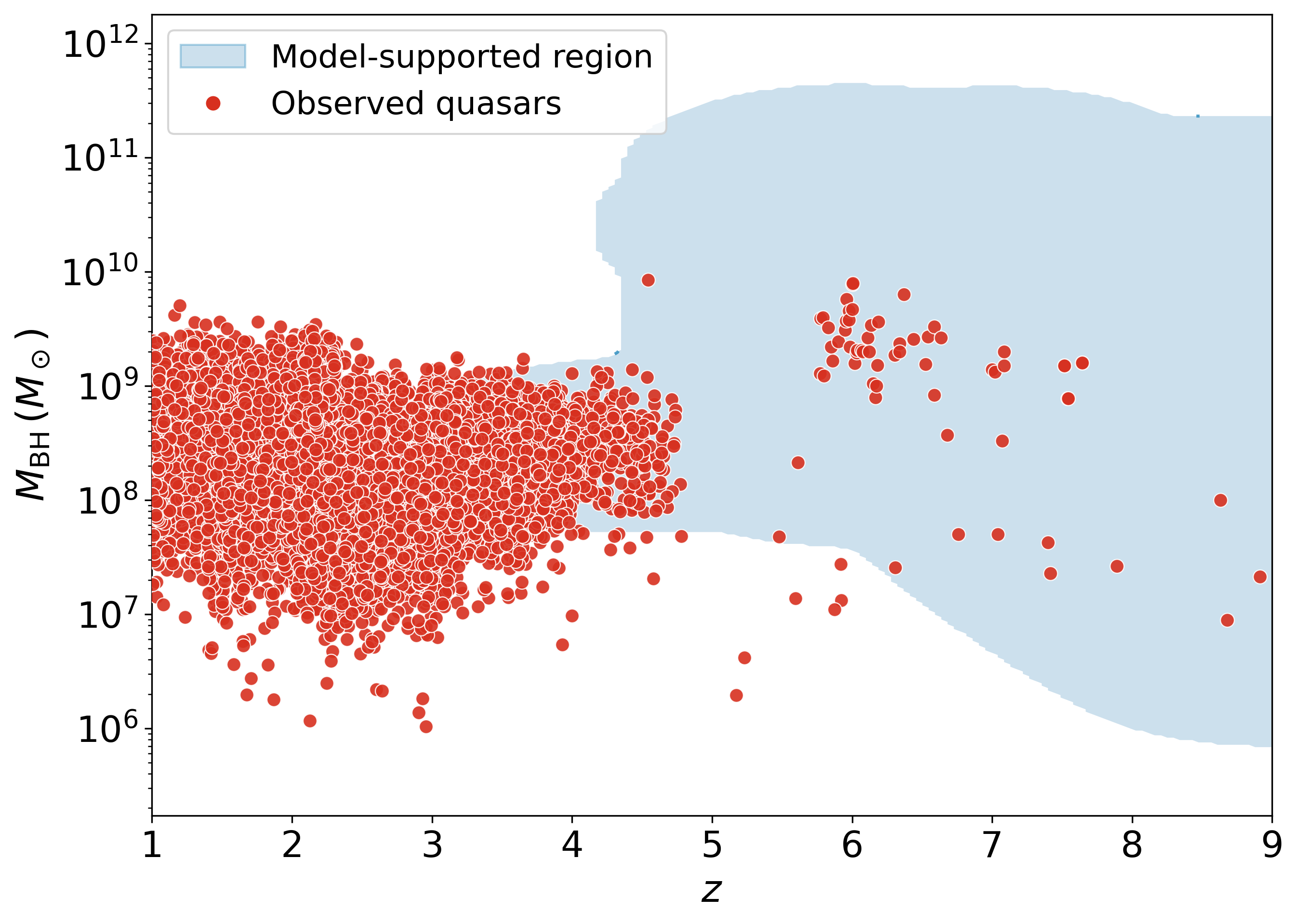}
\caption{Observed quasar sample (red points) compared with the region covered by the model in the $(z,M_{\rm BH})$ plane (blue shaded area). The shaded region is constructed from the complete ensemble of trajectories in the four-dimensional scan, spanning the 4D grid shown at \ref{grid_vals}, with the boson mass sampled at 19 logarithmically spaced values, but in this plot, we just cut the space at $z=9$. We interpolate each trajectory in redshift, and the union of all trajectories defines the blue shaded support region. The figure therefore represents the aggregate coverage of the complete model set rather than a subset corresponding to
a particular value of $\mu$ or $M_{0}$.}
\label{figquasar}
\end{figure}

Figure~\ref{figquasar} shows the observational sample together with the region covered by the complete parameter grid described above. Therefore, the shaded domain in the $(z,M_{\rm BH})$ plane represents the union of all trajectories obtained across the full $(\mu,M_0,M_{c,0},M_{b,0})$ parameter scan, rather than a decomposition into subsets associated with individual values of $\mu$. For visual clarity, the individual numerical trajectories are not displayed as separate curves. Instead, each discrete model evolution is interpolated in redshift, and the shaded region represents the total domain spanned by the ensemble of these interpolated trajectories. In this way, the figure provides a compact visualization of the portion of the observational plane accessible to the model across the explored parameter space. The comparison is intentionally shown in logarithmic mass and linear redshift, so that the full dynamical range of the observed population remains visible within a single plot.

A first visual inspection of Fig.~\ref{figquasar} suggests two qualitative conclusions. On the one hand, the theoretical trajectories cover a substantial portion of the observed SMBH mass scale, including essentially the full high-redshift region with $z>5$. This confirms that, within its effective assumptions, the model is capable of producing very massive BHs at early cosmic times for suitable values of the boson mass and initial conditions. The best-overlap solutions preferentially cluster around relatively light seeds, $M_0\sim10^{2}-10^{3}\,M_{\odot}$, scalar reservoirs of order $M_{c,0}\sim10^{8}-10^{10}\,M_{\odot}$, and baryonic reservoirs in the range $M_{b,0}\sim10^{8}-3\times10^{9}\,M_{\odot}$. The boson mass $\mu$, however, plays a more prominent role in determining the system's overall evolution and is therefore examined in greater statistical detail below. On the other hand, the theoretical domain also extends toward redshifts somewhat larger than those populated by most of the observational sample. This behavior is consistent with the arrival-redshift analysis discussed above: configurations for which the scalar channel operates most efficiently can drive the BH into the SMBH regime significantly earlier, whereas the observed population is more strongly concentrated at lower redshift, where prolonged baryonic accretion may play a more important role. Still, an important portion of the theoretical domain overlaps the observed quasar population for $z\gtrsim3$. This visual comparison, however, is not sufficient by itself to determine which regions of parameter space provide the most representative description of the data, particularly because different trajectories are constrained by different numbers of observational sources.

A quantitative assessment of the compatibility between the model
and the data therefore requires a statistical diagnostic that goes beyond visual inspection. Since the physical quantity of interest spans several orders of magnitude, the natural discrepancy variable is defined in logarithmic mass space. For every numerical run and each observational point that can be consistently compared with that
trajectory, we define the residual
\begin{equation}
\Delta_{i}=
\log_{10}\left(M_{{\rm model},i}\right)
-\log_{10}\left(M_{{\rm obs},i}\right),
\label{eq:delta_i}
\end{equation}
where the numerical mass is evaluated at the redshift corresponding to the observational source. The quality of a given run is then quantified through the global logarithmic root-mean-square error
(RMSE),
\begin{equation}
{\rm RMSE}_{\log M}=
\sqrt{
\frac{1}{N_{\rm comp}}
\sum_{i=1}^{N_{\rm comp}}\Delta_{i}^{2}
},
\label{eq:rmse}
\end{equation}
where $N_{\rm comp}$ is the number of observational sources that can
actually be compared with the corresponding numerical trajectory. This distinction is important in the present analysis, since $N_{\rm comp}$ is not identical for all runs. Complementary metrics such as the mean absolute error,
${\rm MAE}=\langle|\Delta_i|\rangle$, and the median absolute residual may also be employed as consistency diagnostics. Working in $\log_{10}$ mass weights the different mass decades in a balanced manner, which is particularly convenient for a sample extending over several orders of magnitude in $M_{\rm BH}$.

After applying the observational filtering, $\sim 5.000$ models remain in the combined pool; of these, only a subset of about $\sim 875$ has finite global RMSE and enters the quantitative ranking. If these configurations are ranked solely
according to their value of ${\rm RMSE}_{\log M}$, without imposing
a minimum observational coverage, the smallest discrepancy is
\begin{equation}
{\rm RMSE}_{\rm min}=0.1432,
\end{equation}
and corresponds to a boson mass
\begin{equation}
\mu \simeq 1\times10^{-20}\,{\rm eV}.
\label{eq:mu_best_quasar}
\end{equation}
Taken at face value, this result places the best-fitting individual
trajectory in the intermediate-to-high boson-mass sector. However,
this minimum must be interpreted with considerable caution: the
corresponding trajectory is compared with only
$N_{\rm comp}=1$ observational sources. Its very small RMSE therefore
measures an excellent local agreement with a very limited subset of
the catalog rather than a successful global reproduction of the
observed quasar population.

It is consequently useful to distinguish between the quality of the best local match and the statistical support provided by a broad observational coverage. If no minimum value of $N_{\rm comp}$ is imposed and we consider all models lying within
$\Delta{\rm RMSE}\leq0.05$ of the best-fitting configuration, the boson-mass distribution is relatively broad. The corresponding $16$th, $50$th, and $84$th percentiles are
\begin{equation}\begin{split}
&\mu_{16}=3.16\times10^{-21}\,{\rm eV},\\
&
\mu_{50}=1.0\times10^{-20}\,{\rm eV},\\
&
\mu_{84}=5.27\times10^{-20}\,{\rm eV}.
\end{split}
\end{equation}
Thus, when the RMSE alone is used as the ranking criterion, the preferred region shifts toward intermediate and comparatively large boson masses, around $\mu\sim10^{-20}\,{\rm eV}$. Nevertheless, this near-optimal population contains models with very different
observational coverages, including trajectories constrained by only a small number of quasars. The width of this interval should therefore not be interpreted as a conventional statistical confidence interval on $\mu$, but rather as the range sampled by models that provide similarly low residuals under the RMSE-only criterion.

A different and considerably more restrictive picture emerges once observational coverage is explicitly included in the selection. To identify trajectories that describe a non-negligible portion of the
catalog, we require
\begin{equation}
N_{\rm comp}\geq10
\end{equation}
and repeat the near-best selection within this restricted subset. Under this requirement, the favored boson mass becomes concentrated at
\begin{equation}
\mu \simeq 3.162\times10^{-21}\,{\rm eV}\hspace{0.1cm}\rightarrow
\hspace{0.1cm}\mathrm{RMSE}{\min}=0.5123\label{eq:mu_robust_quasar}
\end{equation}
corresponding to one of the discrete values in the numerical $\mu$ grid. In contrast with the absolute RMSE minimum, this result is supported by a significantly broader comparison with the quasar population and should therefore be regarded as the more robust
indicator of the boson-mass scale preferred by the present global analysis. Its precise numerical value is necessarily limited by the discrete resolution of the parameter scan, but the result clearly identifies the ultralight sector around $\mu\sim10^{-21}\,{\rm eV}$
as particularly competitive when observational coverage is taken into account.

The observational analysis reveals a more nuanced picture than the minimum RMSE alone would suggest. Rather than selecting a single sharply preferred boson-mass scale, the results identify a broader phenomenologically relevant interval. Configurations around $\mu\sim10^{-20}\,{\rm eV}$ can yield particularly small individual residuals, although they are generally associated with more limited observational coverage. When a substantial number of quasars is required to enter the comparison, the most robust solutions shift toward lower boson masses, with the preferred value around $\mu\simeq3.16\times10^{-21}\,{\rm eV}$. A conservative interpretation of the present scan therefore places the phenomenologically relevant range approximately between
\begin{equation}
3.16\times10^{-21}\,{\rm eV}
\lesssim \mu
\lesssim
5.27\times10^{-20}\,{\rm eV},
\end{equation}
while recognizing that the distribution within this interval is not uniform and reflects a balance between goodness of fit and observational coverage. Interestingly, this phenomenologically relevant interval overlaps with the boson-mass range already strongly constrained by independent fuzzy-dark-matter probes, including Lyman-$\alpha$ forest measurements and the abundance and structure of Milky Way satellites~\cite{Hui2017,Niemeyer2020,Pozo2025}. The precise interpretation of these bounds is model-dependent, but the overlap highlights the complementarity between early-SMBH growth and small-scale-structure observables in constraining ultralight bosonic dark matter.

This distinction has direct consequences for the physical interpretation of the model. The most efficient scalar-absorption episodes identified in the full parametric study occur approximately in the range
\begin{equation}
\mu\sim10^{-19}\text{-}10^{-16}{\rm eV}.
\end{equation}
The observational comparison therefore does not indicate a complete separation between quasar-compatible configurations and the efficient scalar-growth regime. The higher-$\mu$ tail of the observationally competitive models approaches the lower boundary of the efficient-absorption window, showing that some configurations with a non-negligible scalar contribution can also reproduce selected portions of the observed quasar distribution particularly well. However, the most strongly scalar-dominated solutions are not, in general, those supported by the broadest observational coverage. This suggests that maximizing the efficiency of scalar-assisted BH growth and providing a representative description of the observed quasar population are related, but distinct, requirements.

When broad observational coverage is demanded, the preferred value instead remains in the ultralight sector, $\mu\simeq3.16\times10^{-21}\,{\rm eV}$, well below the region of
maximum scalar-absorption efficiency. For such masses, the dramatic early scalar-growth mechanism is comparatively weak in the present model, and the baryonic component more strongly shapes the evolutionary tracks. The fact that these trajectories provide the most stable agreement with a wider portion of the catalog indicates that reproducing the bulk of the observed quasar population and producing the fastest possible early SMBH growth are not necessarily the same requirement. In this sense, the two statistical modes found here may reflect two different phenomenological regimes of the model rather than a single narrowly localized preferred boson mass.

Interestingly, the lower mass scale is also close to the ultralight boson masses frequently considered in scenarios in which scalar fields contribute to galactic or intragalactic dark-matter phenomenology. Conversely, the higher-mass mode lies much closer to the parameter region in which scalar absorption can rapidly increase the BH mass. One possible interpretation is therefore that a single scalar mass may not simultaneously maximize the global agreement with the quasar population and the efficiency of the earliest SMBH-growth episodes. Alternatively, a multi-mass scalar scenario could accommodate both behaviors, with an ultralight component associated with the broadly quasar-compatible evolutionary tracks and a heavier component capable of triggering rapid high-redshift absorption. Such an interpretation remains speculative and is not uniquely selected by the current dataset; importantly, the updated results also leave open the simpler possibility that a single intermediate mass could provide an acceptable compromise once the parameter space is explored more densely.

Finally, several caveats intrinsic to the mass observational comparison deserve to be stated explicitly. The quasar catalog is subject to selection effects, including the preferential detection of luminous, actively accreting sources and the increasingly
incomplete coverage of the lower-mass population toward high redshift. Virial and single-epoch BH mass determinations additionally carry substantial statistical and systematic uncertainties that are not explicitly folded into the simple residual statistic of
Eq.~(\ref{eq:delta_i}). On the theoretical side, the numerical trajectories are compared only through the BH mass and redshift, without incorporating luminosity, duty-cycle information, or a survey selection function, all of which would be required for a fully self-consistent likelihood analysis against flux-limited catalogs.

A further limitation, made particularly clear by the present updated analysis, is that the RMSE and observational coverage contain complementary information. An exceptionally small RMSE obtained from only a few objects cannot be assigned the same statistical weight as a moderately larger residual obtained across a substantial fraction of the available comparison points. The numerical values quoted above should therefore be viewed as diagnostics of the present parameter scan rather than as formal posterior constraints on $\mu$. Moreover, the discrete sampling of the boson mass implies that the concentration at $\mu=3.162\times10^{-21}\,{\rm eV}$ is resolved only at the current grid spacing, while a denser exploration around both the low- and high-$\mu$ modes may shift their precise locations. These limitations motivate a natural next step: incorporating goodness of fit and observational coverage simultaneously into a weighted or likelihood-based statistic, ultimately supplemented by the observational uncertainties and selection function of a homogeneous quasar catalog. Such an analysis will determine whether the apparent bimodal structure in $\mu$ persists or whether one regime becomes statistically preferred once the comparison is made on equal observational footing.

% ---------------------------------
\subsection{Luminosity imprint of scalar-assisted growth}
\label{sec:luminosity}
% ---------------------------------

The comparison presented above tests whether the numerical growth trajectories can reproduce the observed distribution of quasar BH masses in the $(z,M_{\rm BH})$ plane. However, the presence of the scalar component may leave an additional observational imprint through the baryonic accretion history. Since the scalar reservoir contributes directly to the growth of the BH, the resulting increase in $M_{\rm BH}$ modifies the subsequent Eddington-like baryonic accretion rate through Eq.~(\ref{eq:medd}). The scalar channel can therefore affect not only the epoch at which a given BH mass is reached, but also the subsequent evolution of the luminosity associated with baryonic accretion.

To quantify this effect, we reconstruct the baryonic contribution tothe BH growth directly from the evolution of the baryonic reservoir.
Using Eq.~(\ref{eq:Mbdot}), we define
\begin{equation}
\dot M_{{\rm BH},b}(z)
=
\max\left[
-\frac{dM_b}{dt},
0
\right],
\label{eq:Mdot_b_lum}
\end{equation}
which coincides with the effective baryonic mass-transfer rate whenever the reservoir is being depleted. The corresponding bolometric luminosity proxy is written as
\begin{equation}
L_{\rm bar}(z)
=
\frac{\epsilon}{1-\epsilon}
\dot M_{{\rm BH},b}(z)c^2,
\label{eq:Lbar}
\end{equation}
where the factor $\epsilon/(1-\epsilon)$ follows from the convention adopted in Eq.~(\ref{eq:medd}), in which $\dot M_{{\rm BH},b}$ denotes the net mass added to the BH after radiative losses. For comparison with the observational quasar sample, we use $L_{\rm bar}$ as a proxy for the bolometric luminosity and consider
\begin{equation}
\log_{10}
\left(
\frac{L_{\rm bar}}
{{\rm erg s^{-1}}}
\right).
\label{eq:logLbar}
\end{equation}

It is important to emphasize that the scalar field does not contribute directly to Eq.~(\ref{eq:Lbar}) as an additional radiative component. Its effect is indirect: scalar absorption increases $M_{\rm BH}$, thereby modifying the subsequent baryonic Eddington-like feeding rate and, consequently, the luminosity associated with ordinary matter accretion. The luminosity reconstruction therefore provides a complementary probe of the dynamical feedback already identified in the mass evolution.

To isolate this effect, each evolution including scalar-field accretion is compared with a corresponding control trajectory in which the scalar contribution to BH growth is absent, while all other initial conditions and model ingredients are kept unchanged. The control trajectories therefore describe the corresponding evolution under baryonic accretion alone. We first quantify the differences between each pair of trajectories before comparing them with observations.

A useful measure of the cumulative effect is the radiative output
integrated over the redshift interval of interest,
\begin{equation}
E
=
\int_{z_{\min}}^{z_{\max}}
L_{\rm bar}(z)
\left|
\frac{dt}{dz}
\right|dz,
\label{eq:Erad}
\end{equation}
where $dt/dz$ is given by Eq.~(\ref{eq:dtdz}). We then define
\begin{equation}
\mathcal{R}_{E}
\equiv
\frac{E_{\rm with}}{E_{\rm control}},
\label{eq:energy_ratio}
\end{equation}
where, throughout this analysis, the labels ``with'' and ``control'' refer, respectively, to the scalar-assisted evolution and to the corresponding evolution in which the scalar contribution is switched off, while keeping the remaining initial conditions unchanged. Thus, $E_{\rm with}$ and $E_{\rm control}$ are obtained by evaluating Eq.~(\ref{eq:Erad}) using the baryonic luminosities $L_{\rm bar}^{\rm with}(z)$ and $L_{\rm bar}^{\rm control}(z)$, respectively, such that $\mathcal{R}_{E}>1$ indicates a larger integrated baryonic radiative output in the scalar-assisted case. The same ``with'' and ``control'' notation will be used below for other quantities whenever the scalar-assisted and reference evolutions are compared.

A complementary diagnostic is obtained from the depletion timescale of the baryonic reservoir. Defining $z_{50}$ through
\begin{equation}
M_b(z_{50})
=
\frac{1}{2}M_{b,0},
\end{equation}
we introduce
\begin{equation}
\Delta z_{50}
=
z_{50}^{\rm with}
-
z_{50}^{\rm control}.
\label{eq:dz50}
\end{equation}

Because larger redshifts correspond to earlier cosmic time, $\Delta z_{50}>0$ implies that the baryonic reservoir reaches half of its initial mass earlier when the scalar contribution is present. Unlike the integrated luminosity ratio, this quantity is primarily sensitive to a redistribution of the accretion history over time.

\begin{figure}[t]
\centering
\includegraphics[width=\columnwidth]{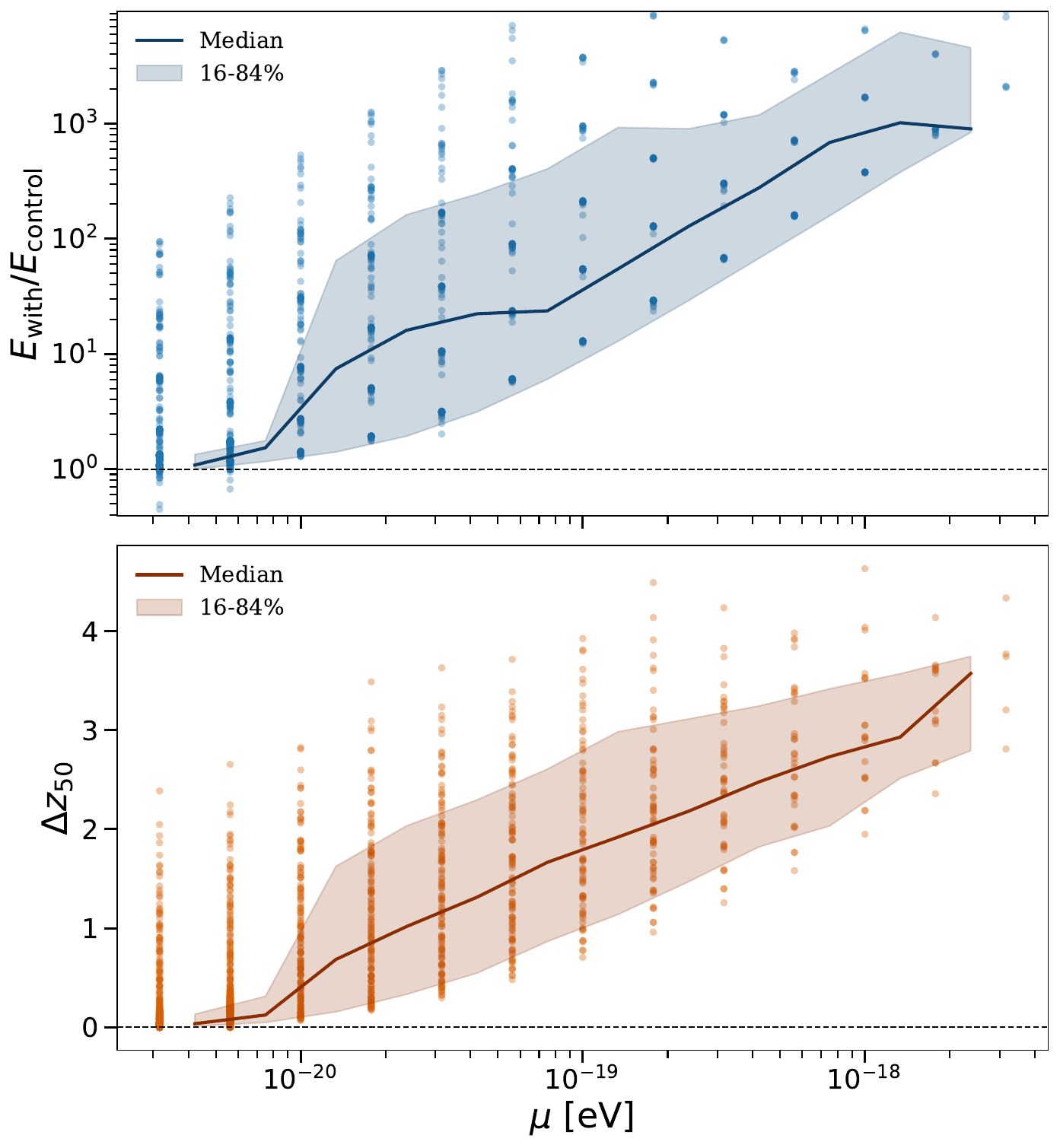}
\caption{Scalar-impact diagnostics as a function of the boson mass $\mu$. The upper panel shows the integrated luminosity ratio between the scalar-assisted and control realizations, while the lower panel displays the corresponding shift in the baryonic-depletion epoch. Individual points represent paired model realizations, whereas the solid curves and shaded regions indicate the median and the $16$th--$84$th percentile interval across the ensemble.}
\label{fig:scalar_impact}
\end{figure}

Figure~\ref{fig:scalar_impact} summarizes these two intrinsic diagnostics as a function of the boson mass. The upper panel shows the integrated luminosity ratio, while the lower panel displays the corresponding shift in the baryonic-depletion epoch. Individual points retain the variability among paired realizations with different initial conditions, whereas the median curves and percentile bands highlight the systematic dependence on $\mu$.

These quantities characterize the dynamical impact of the scalar channel independently of any direct comparison with the observational normalization. In particular, they make it possible to distinguish between an enhancement of the cumulative baryonic radiative output and a shift in the epoch at which that output is produced. This distinction is relevant because scalar-assisted growth need not simply increase the luminosity uniformly: it may instead alter the temporal distribution of the baryonic accretion history.

\begin{figure}[t]
\centering
\includegraphics[width=\columnwidth]{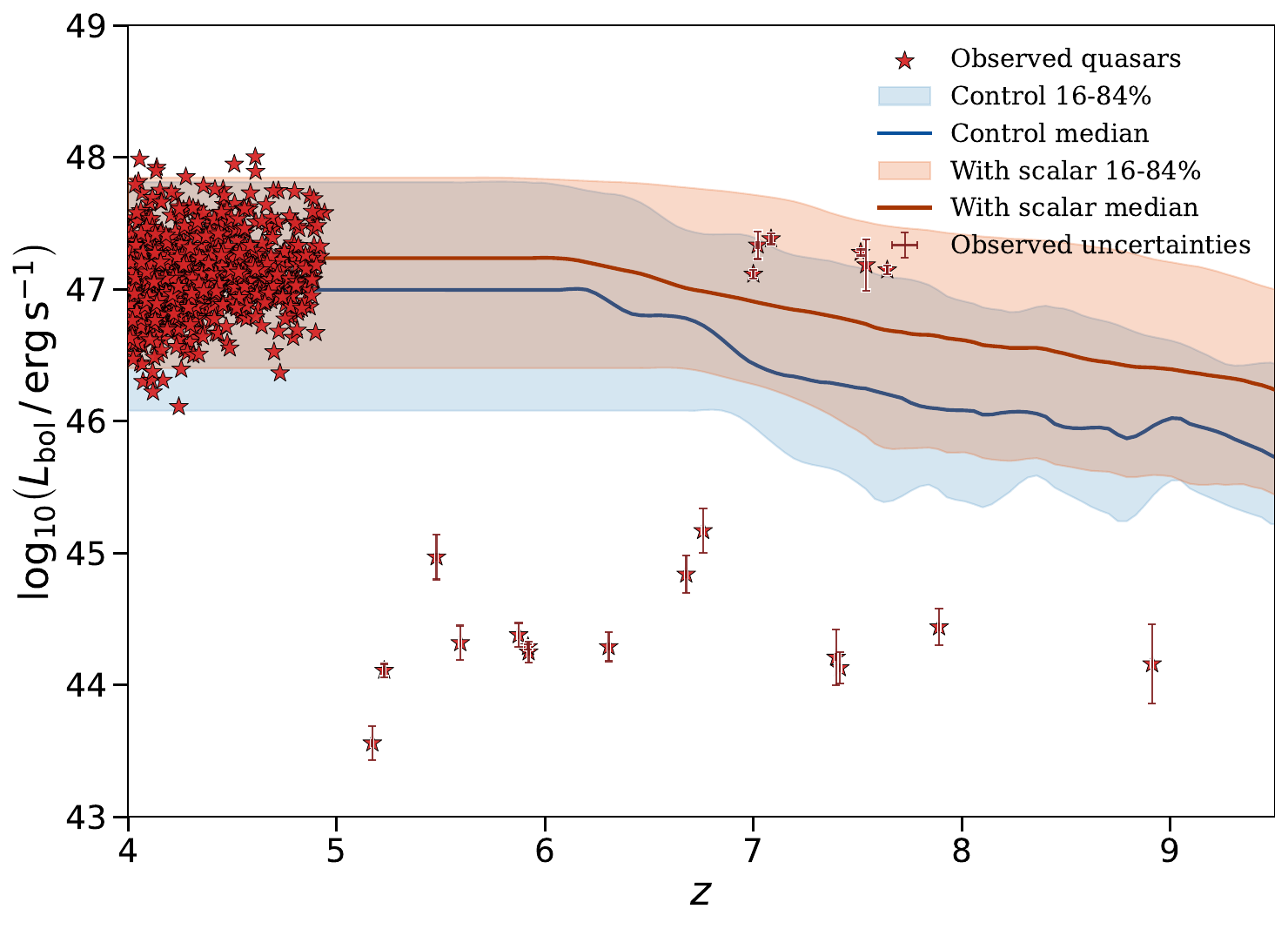}
\caption{Redshift evolution of the bolometric luminosity,
$\log_{10}(L_{\rm bol}/{\rm erg,s^{-1}})$, over the interval
$4\leq z\leq9.5$. Red stars denote observed high-redshift quasars and AGN from the luminosity subsample of the merged catalog (restricted to $4\leq z\leq 9$ and requiring available luminosity-uncertainty estimates), with error bars showing the reported uncertainties.. The blue solid line and
shaded region correspond to the control scenario, while the orange solid line and shaded region represent the scalar-assisted case; in both cases, the solid curve denotes the median and the shaded band the $16$th--$84$th percentile interval across the model ensemble. Model luminosities are obtained from the baryonic-accretion proxy and calibrated through a global additive shift in $\log L$ determined over $4\leq z\leq5$. The comparison therefore emphasizes the redshift-dependent evolution, the relative displacement between the control and scalar scenarios, and their overlap with the observed luminosity distribution.}
\label{fig:lbol_evolution}
\end{figure}

We next confront the reconstructed luminosity histories directly with the observational sample. Figure~\ref{fig:lbol_evolution} shows the population-level evolution in the $\log L_{\rm bol}$--$z$ plane over the interval $4\leq z\leq9.5$. For both the control and scalar-assisted ensembles, the solid curve represents the median luminosity at each redshift, while the shaded region denotes the $16$th--$84$th percentile interval. The observed high-redshift quasars and AGN are superimposed for comparison.

Before performing this comparison, the theoretical luminosity proxy is calibrated through a single global additive shift in $\log L$, determined within the reference interval $4\leq z\leq5$. This calibration removes an overall normalization offset and allows the comparison to focus on the redshift dependence of the luminosity evolution. Consequently, the relevant information is not an absolute vertical displacement between the theoretical and observational distributions, but rather the relative evolution of the control and scalar-assisted populations, their separation with increasing redshift, and their overlap with the observed luminosity distribution.

A systematic separation between the two theoretical populations in
Fig.~\ref{fig:lbol_evolution} would therefore constitute a direct
signature of the influence of scalar-assisted growth on the baryonic
luminosity history, even though the scalar transfer itself is not
assumed to radiate in the present framework. In particular, an upward
shift of the scalar-assisted median relative to the control case would
indicate an enhancement of the effective baryonic luminosity, whereas
changes in the shape or redshift dependence of the two curves would
signal a redistribution of the accretion history. The width of the
percentile bands additionally reflects how strongly this response
depends on the remaining parameters of the model.

The visual comparison alone, however, is not sufficient to determine
whether the scalar contribution provides a systematically better
description of the observational trend. We therefore introduce an
uncertainty-weighted residual statistic. For each value of $\mu$, the
model realizations are grouped together and used to construct the
median scalar-assisted and control luminosity curves. These are then
compared with the observational $\log L_{\rm bol}(z)$ relation binned
in redshift. For either scenario, we define
\begin{equation}
{\rm WRMSE}
=
\left[
\frac{1}{N}
\sum_{i=1}^{N}
\left(
\frac{m_i-o_i}{\sigma_i}
\right)^2
\right]^{1/2},
\label{eq:wrmse_lum}
\end{equation}
where $m_i$ is the model prediction in the $i$th redshift bin, $o_i$
is the corresponding observational median, $\sigma_i$ is the effective
observational uncertainty, and $N$ is the number of valid bins entering
the comparison.

To isolate the relative performance of the two dynamical scenarios,
we then define
\begin{equation}
\Delta{\rm WRMSE}(\mu)
=
{\rm WRMSE}_{\rm with}(\mu)
-
{\rm WRMSE}_{\rm control}(\mu).
\label{eq:delta_wrmse}
\end{equation}
With this convention, $\Delta{\rm WRMSE}<0$ indicates that the scalar-assisted population follows the observational luminosity trend more closely than the corresponding control population, while $\Delta{\rm WRMSE}>0$ favors the control scenario. The $\Delta{\rm WRMSE}=0$ line therefore represents equal performance.

\begin{figure}[t]
\centering
\includegraphics[width=\columnwidth]{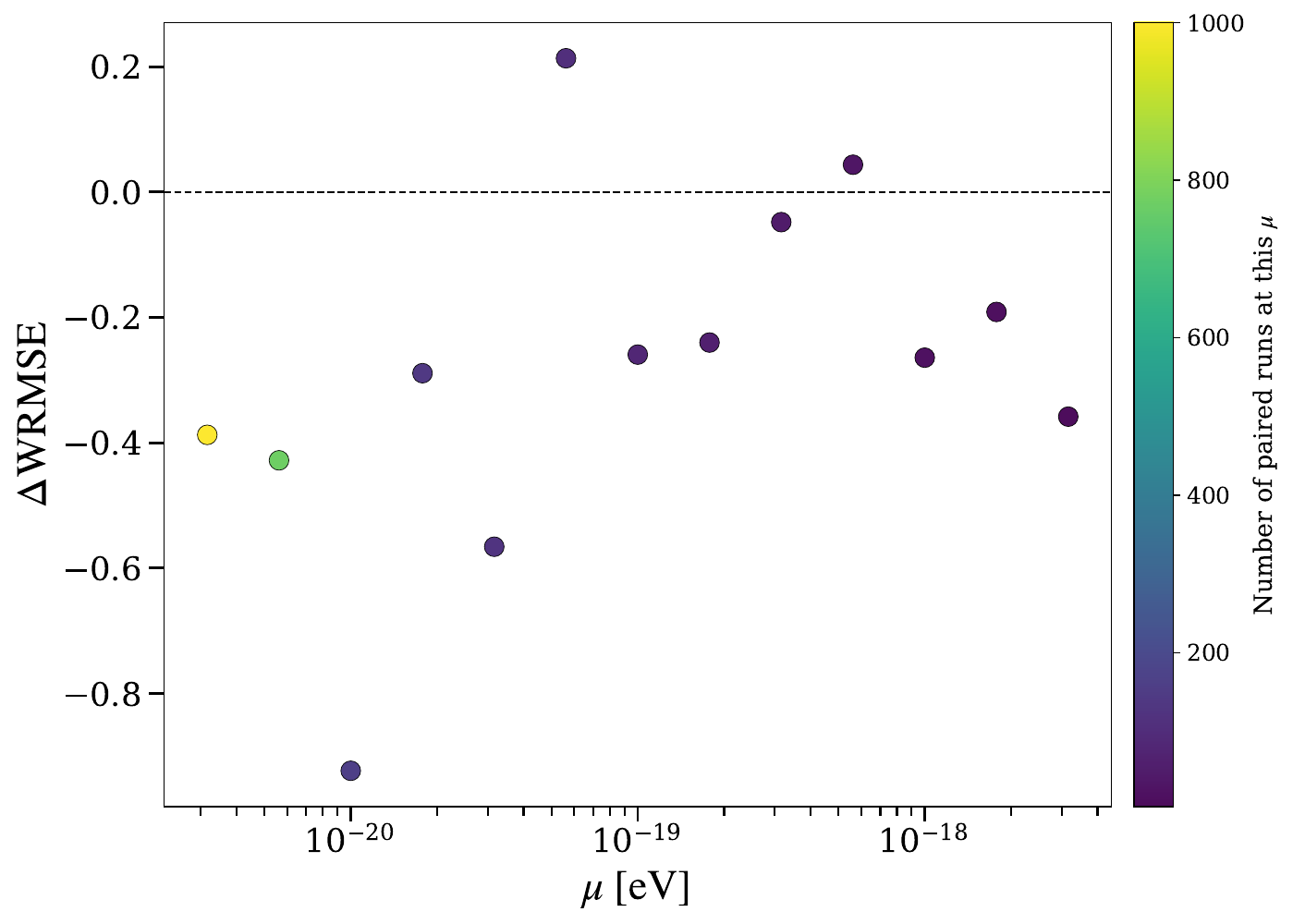}
\caption{Difference in the observationally weighted mismatch between the scalar-assisted and control scenarios as a function of the boson mass $\mu$. Each WRMSE is computed with respect to the binned observational $\log L_{\rm bol}(z)$ relation over the adopted redshift interval, with residuals weighted by the corresponding observational uncertainty in each bin. Negative values of $\Delta{\rm WRMSE}$ favor the scalar-assisted model, while positive values favor the control scenario; the horizontal line at $\Delta{\rm WRMSE}=0$ marks equal performance. Marker color encodes the number of paired model realizations contributing to each value of $\mu$, indicating the statistical support of the comparison.}
\label{fig:delta_wrmse}
\end{figure}

Figure~\ref{fig:delta_wrmse} displays this differential statistic as afunction of the boson mass. In addition to the value of $\Delta{\rm WRMSE}$, the marker color encodes the number of paired realizations contributing to each value of $\mu$. This information is important when assessing the robustness of individual points, since comparisons supported by a larger number of model pairs provide a more representative estimate of the behavior of that region of parameter space.

The differential comparison identifies a clear region where the scalar-assisted models better describe the observed luminosity trend than their control counterparts. The strongest relative improvement occurs at
\begin{equation}
\mu_{\star\Delta}=1.0\times10^{-20}\,{\rm eV},
\end{equation}
for which $\Delta{\rm WRMSE}=-0.9232$, based on $n_{\rm pairs}=160$ paired realizations. If instead the scalar-assisted models are ranked by their absolute agreement with the observations, the minimum value of ${\rm WRMSE}_{\rm with}$ is obtained at
\begin{equation}
\mu_{\star{\rm abs}}=5.623\times10^{-21}\,{\rm eV},
\end{equation}
with ${\rm WRMSE}_{\rm with}=0.2059$ and $n_{\rm pairs}=768$. The two criteria therefore select nearby, but not identical, boson-mass scales, with the preferred values lying around a few $\times10^{-21}$-$10^{-20}{\rm eV}$.

A more robust interval is obtained by simultaneously requiring an actual improvement over the control case, $\Delta{\rm WRMSE}<0$, and a substantial number of paired realizations, $n_{\rm pairs}\geq100$. Under these conditions, the favored region extends approximately over
\begin{equation}
3.16\times10^{-21}\,{\rm eV}
\lesssim
\mu
\lesssim
3.16\times10^{-20}\,{\rm eV}.
\end{equation}
Weighting the improving bins by their observational coverage yields a characteristic value
\begin{equation}
\mu_{50,w}\simeq3.83\times10^{-21}\,{\rm eV},
\end{equation}
with a corresponding weighted interval
\begin{equation}
\mu_{16,w}\simeq3.16\times10^{-21}\,{\rm eV},
\qquad
\mu_{84,w}\simeq1.58\times10^{-20}\,{\rm eV}.
\end{equation}
These results indicate that the luminosity comparison consistently favors an ultralight-to-intermediate boson-mass range, with the strongest observational support concentrated around $\mu\sim10^{-21}$-$10^{-20}\,{\rm eV}$.

The three diagnostics considered in this subsection therefore probe distinct but complementary consequences of the same scalar-induced feedback. The quantities $\mathcal{R}_{E}$ and $\Delta z_{50}$ measure how the scalar channel changes the intrinsic baryonic accretion history, respectively through its cumulative radiative output and its characteristic depletion timescale. Figure~\ref{fig:lbol_evolution} then shows how these modifications affect the population-level luminosity evolution, while $\Delta{\rm WRMSE}$ provides a direct measure of whether the resulting redshift dependence is more or less consistent with the observed quasar population.

This distinction is important because a strong dynamical effect does not necessarily lead to better agreement with observations. Conversely, even a modest scalar contribution may change the luminosity evolution in a way that improves the agreement with the data. We therefore compare the dynamical and observational effects as a function of $\mu$. Figures~\ref{fig:scalar_impact}, \ref{fig:lbol_evolution}, and~\ref{fig:delta_wrmse} show whether scalar-assisted BH growth produces both a measurable dynamical effect and a corresponding change in the high-redshift quasar luminosity distribution.

Finally, the luminosity defined in Eq.~(\ref{eq:Lbar}) should be interpreted as an effective proxy for the baryonic depletion history. It does not model the full radiative physics of quasar accretion and does not account for spectral corrections, obscuration, anisotropic emission, or time-dependent duty cycles. The luminosity analysis should therefore be viewed as a differential comparison between scalar-field and control growth histories, rather than as a complete model of the observed quasar population. Within these limitations, the luminosity analysis tests whether scalar-induced changes in BH growth lead to systematic changes in the baryonic accretion history that may be reflected in the observations.

%----
\section{Conclusions}\label{sec:conclusions}
%----

We present a phenomenological study of cosmological BH growth in which a finite baryonic reservoir and a bosonic scalar cloud contribute simultaneously to the evolution of BH mass. The system is evolved directly in redshift within a flat $\Lambda$CDM background, and its dependence on the boson mass $\mu$, the initial seed mass $M_{0}$, the initial scalar-cloud mass $M_{c,0}$, and the initial baryonic reservoir mass $M_{b,0}$ is systematically explored through a broad multidimensional parameter scan. The framework combines an Eddington-like prescription for the baryonic sector with an effective quasibound-state decay rate for the scalar channel, allowing the competition between the two growth mechanisms to be followed consistently throughout cosmological evolution.

The scalar contribution to BH growth is strongly dependent on the boson mass, the BH mass, and the initial scalar-cloud mass. Efficient scalar absorption occurs only in a restricted region of parameter space, with the relevant scale set by the dimensionless coupling $\alpha=\mu M_{\rm BH}$. For the SMBH masses considered here, the most efficient growth occurs approximately for $\mu\sim10^{-19}$--$10^{-16}\,\mathrm{eV}$. When a sufficiently massive scalar reservoir is present, absorption can substantially accelerate BH growth and, in favorable cases, advance the formation of $10^9M_\odot$ BHs to higher redshifts. In the most extreme cases, scalar accretion can temporarily dominate over baryonic growth during the early stages of SMBH assembly.
The observational comparisons, however, favor a lower boson-mass range. The BH-mass analysis identifies near-optimal solutions predominantly around $\mu\sim10^{-21}$--$10^{-20}\,\mathrm{eV}$, while the luminosity analysis leads to a similar range. In the latter case, the scalar field affects the luminosity indirectly by changing the BH mass and therefore the subsequent baryonic accretion history. Thus, the boson mass that maximizes the dynamical impact of scalar accretion is not necessarily the one that gives the best agreement with the observed quasar population. Rather, the results indicate that moderate scalar contributions at lower $\mu$ can provide better observational agreement than the strongest scalar-growth scenarios.
These results also show that the initial scalar-cloud mass plays an important role in determining the magnitude of the effect, while the boson mass primarily controls whether scalar transfer becomes efficient. The observationally favored and dynamically efficient regimes are therefore distinct, although they are not necessarily unrelated. A more detailed treatment of the scalar-cloud formation and evolution, together with a more realistic description of BH accretion and quasar luminosities, will be needed to determine whether an intermediate region can simultaneously produce substantial scalar-driven growth and good agreement with the observations.
The mass- and luminosity-based comparisons point toward a similar observationally favored range, approximately $\mu\sim10^{-21}$--$10^{-20}\,\mathrm{eV}$, with some solutions extending to a few $10^{-20}\,\mathrm{eV}$. The two diagnostics probe different aspects of the evolution: the mass comparison tests the BH growth trajectories in the $(z,M_{\rm BH})$ plane, while the luminosity comparison probes the effect of scalar accretion on the subsequent baryonic accretion history. Their agreement therefore suggests that the preference for lower boson masses is not specific to a single observable. Importantly, the mass range that maximizes scalar-growth efficiency, $\mu\sim10^{-19}$--$10^{-16}\,\mathrm{eV}$ for the SMBH masses considered here, is different from the range favored by the observations. This distinction follows from the dependence of the scalar-growth efficiency on $\alpha=\mu M_{\rm BH}$ and from the fact that the observational comparisons probe different stages of the BH evolution.
The present results should nevertheless be interpreted within the limitations of the adopted model. The scalar transfer rate is described by an effective small-coupling prescription and does not include the full scalar-cloud profile, self-gravity, formation history, or higher multipoles. The BH is also assumed to be non-rotating, while the baryonic sector is described by an Eddington-like prescription with fixed radiative efficiency. Realistic SMBH growth may additionally involve super-Eddington accretion, radiative feedback, mergers, evolving gas supplies, and time-dependent duty cycles. The observational analysis is likewise phenomenological: the mass comparison does not constitute a formal likelihood analysis, while the luminosity is reconstructed from the baryonic depletion history rather than obtained from a complete radiative model. The preferred boson-mass ranges should therefore be regarded as results of the present framework rather than as formal constraints on $\mu$.
Future work should address these limitations by including scalar-field backreaction and BH spin evolution, together with a more realistic treatment of baryonic accretion and quasar emission. A denser exploration of the intermediate boson-mass range will also be useful for determining whether a single boson mass can simultaneously produce appreciable scalar-driven growth and good agreement with the observations. More generally, embedding the model in cosmological structure-formation calculations would provide physically motivated initial conditions for the scalar and baryonic reservoirs and allow the role of scalar accretion to be assessed in a more realistic evolutionary setting.

\begin{acknowledgments}
JCM is funded by a UNAM-DGAPA Postdoctoral Fellowship and by the PAPIIT-IN108526 Determinación de consecuencias observacionales de la naturaleza
de las componentes obscuras del Universo project. SF is funded by the Conselleria de Innovación, Universidades, Ciencia y Sociedad Digital of the Generalitat Valenciana and the European Social Fund through a postdoctoral fellowship APOSTD 2025 (CIAPOS/2024/461). NSG acknowledges support from the Spanish Ministry of Science, Innovation, and Universities via the Ram\'on y Cajal programme (grant RYC2022-037424-I), funded by  MICIU/AEI/10.13039/501100011033 and by ESF+. This work is also supported by the Spanish Agencia Estatal de Investigación (grant PID2024-159689NB-C21) funded by MICIU/AEI/10.13039/501100011033 and by FEDER/EU, by the Generalitat Valenciana (Prometeo grant CIPROM/2022/49), by PAPIIT grant IN107026, and by the European Horizon Europe staff exchange (SE) programme HORIZON-MSCA2021-SE-01 Grant No. NewFunFiCO-101086251.
\end{acknowledgments}

%--
\bibliographystyle{apsrev4-2}
\bibliography{biblio}
%--

\appendix

% -----------------------------------------------------------------
\section{Bosonic clouds at high redshifts}
\label{app:clouds}
% -----------------------------------------------------------------

A gravitationally bound scalar cloud can form at \(z>120\) without violating any universal cosmological bound. For a scalar field with constant mass \(\mu\) and an approximately quadratic potential, coherent oscillations begin when \(3H(z_{\rm osc})\simeq\mu\), after which its mean energy density evolves as nonrelativistic matter. In units \(c=\hbar=1\), during radiation domination this condition yields \(1+z_{\rm osc}\simeq1.55\times10^{6}(\mu/10^{-22}\,\mathrm{eV})^{1/2}\); hence, even for \(\mu=10^{-24}\,\mathrm{eV}\), the oscillatory regime begins well before \(z=120\) \cite{Diez-Tejedor:2017ivd}. However, the onset of oscillations does not by itself imply the formation of a bound structure. To illustrate the conditions for collapse, consider the nonrelativistic limit with negligible self-interactions, assuming that the scalar field dominates the clustering matter component. Its linear density perturbations satisfy
\begin{equation}
\begin{split}
&\ddot{\delta}_k+2H\dot{\delta}_k+
\left(\frac{k^4}{4\mu^2a^4}-4\pi G\bar{\rho}_{m}\right)\delta_k=0,
\\&
k_J(a)=a\left(16\pi G\bar{\rho}_{m}\mu^2\right)^{1/4},
\end{split}
\end{equation}
where \(k\) denotes the comoving wavenumber \cite{Hu2000}. On scales sufficiently larger than the Jeans scale, \(k\ll k_J\), gradient pressure becomes negligible and the growing mode obeys \(\delta_k\propto a\) during matter domination. Within the spherical-collapse approximation, a perturbation can collapse at \(z_{\rm col}>120\) provided that
\[
\delta_L(k,z_i)\,
\frac{D(z_{\rm col})}{D(z_i)}
\simeq\delta_c,\qquad
\delta_c\simeq1.686,
\]
where \(D(z)\) is the linear growth factor and \(\delta_L\) is the linearly extrapolated overdensity. As an illustrative estimate, adopting \(D\propto(1+z)^{-1}\) from matter–radiation equality, \(z_{\rm eq}\simeq3400\), collapse at \(z_{\rm col}=120\) requires an amplitude of order \(\delta_{L,\rm eq}\sim1.686(121/3401)\simeq0.06\). This estimate is approximate, since an accurate calculation must account for the transition between radiation and matter domination. Thus, sufficiently enhanced perturbations on scales unaffected by gradient-pressure suppression can collapse before \(z=120\). Postinflationary axion-like scenarios provide concrete mechanisms for generating isocurvature perturbations that enhance early structure formation \cite{Irsic:2019iff}, while simulations of axion miniclusters demonstrate that collapse can even begin during radiation domination \cite{Eggemeier:2019khm}. These results establish the possibility of this mechanism, but do not demonstrate its viability for arbitrary scalar masses or abundances. Consequently, a bound scalar cloud at \(z>120\) is a physically admissible initial condition only if the production scenario satisfies the growth and Jeans conditions, together with the relevant observational constraints. If a relaxed solitonic core is additionally assumed, its formation timescale must also fit within the available cosmic time \cite{Levkov:2018kau}.

% -----------------------------------------------------------------
\section{Additional redshift-evolution plots}
\label{app:trajplots}
% -----------------------------------------------------------------

In this appendix, we collect the representative redshift-evolution
diagrams corresponding to the three values of the initial
scalar-cloud mass $M_{c,0}$ that were omitted from the main text
for reasons of space. Together with the reference case
$M_{c,0}=10^{9}\,M_\odot$ discussed in
Sec.~\ref{sec:results}, the four values
$M_{c,0}\in\{10^{7},10^{8},10^{9},10^{10}\}\,M_\odot$ span the
entire range of scalar reservoirs sampled in the parameter scan
and allow the dependence of the coupled evolution on the initial
cloud mass to be visualized in a controlled and homogeneous way.
In every panel presented below, the initial BH seed mass is fixed
to $M_{0}=10^{3}\,M_\odot$ and the initial baryonic reservoir is
fixed to $M_{b,0}=5\times 10^{9}\,M_\odot$, so that the four
values of $M_{c,0}$ correspond to progressively larger scalar
components relative to a common baryonic background. The four
representative boson masses shown in each panel,
$\mu\in\{10^{-16},10^{-17},10^{-18},10^{-19}\}\,{\rm eV}$, cover
the interval in which the scalar channel transitions from strongly
efficient to essentially inactive within the redshift window
considered in this work. As throughout the paper, solid lines
represent the BH mass $M_{\rm BH}(z)$, dashed lines the mass
$M_{c}(z)$ remaining in the scalar cloud, and dotted lines the
mass $M_{b}(z)$ still available in the baryonic reservoir; the
color of each triplet identifies the corresponding value of $\mu$.

\begin{figure}[t]
    \centering
    \includegraphics[width=\columnwidth]{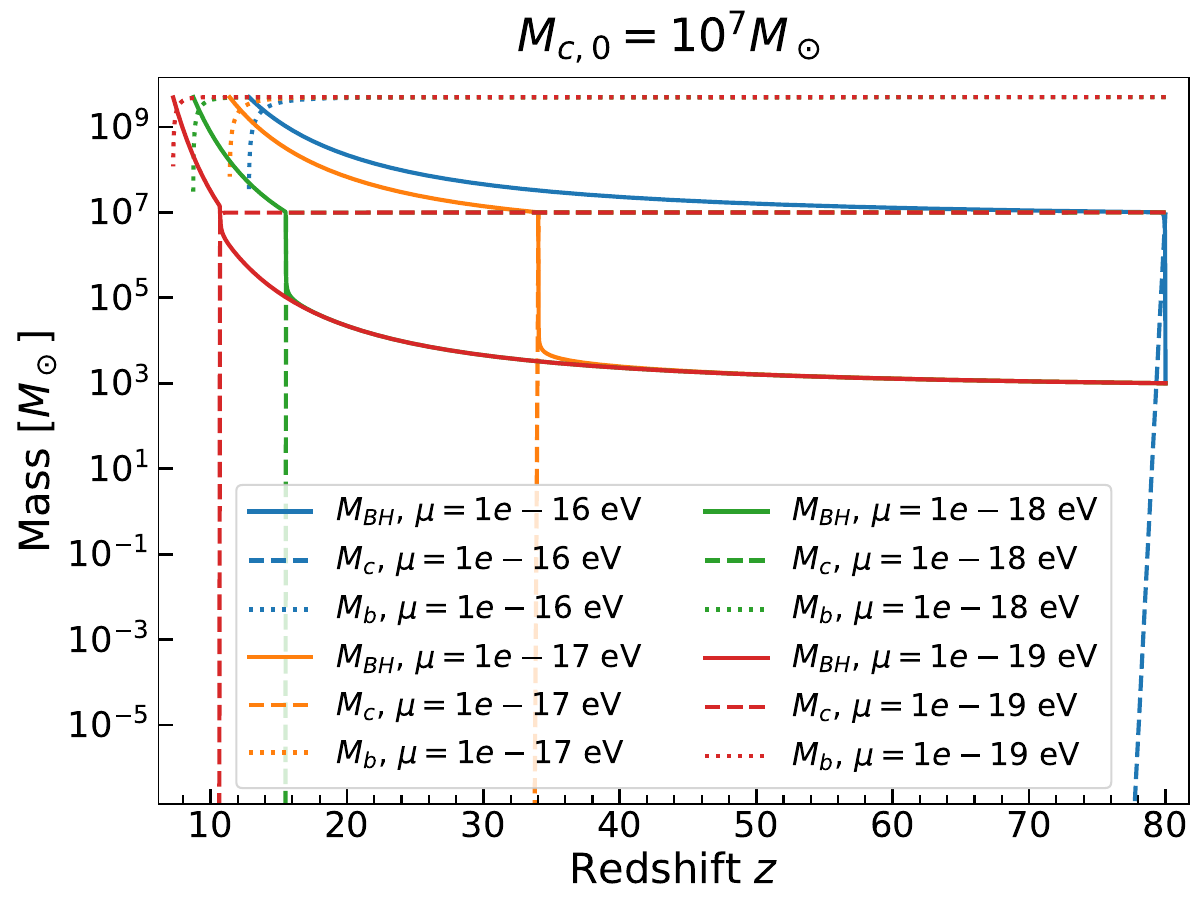}
    \caption{Redshift evolution of $M_{\rm BH}$ (solid), $M_{c}$
    (dashed), and $M_{b}$ (dotted) for $M_{c,0}=10^{7}\,M_\odot$,
    with fixed $M_{0}=10^{3}\,M_\odot$ and
    $M_{b,0}=5\times 10^{9}\,M_\odot$, for four representative
    values of the boson mass $\mu$.}
    \label{fig:evol_mc1e7}
\end{figure}

Figure~\ref{fig:evol_mc1e7} shows the evolution of the three mass
components for the smallest scalar reservoir considered here,
$M_{c,0}=10^{7}\,M_\odot$. In this configuration, the scalar
component is about two orders of magnitude below the baryonic
reservoir, so that the cloud cannot on its own drive the BH into
the SMBH regime. Nevertheless, the four values of $\mu$ produce
very different dynamical behaviors. For $\mu=10^{-16}\,{\rm eV}$
the scalar-decay rate is so large that the cloud is essentially
absorbed at the very beginning of the integration, near
$z\simeq 80$: the dashed blue curve drops abruptly to values many
orders of magnitude below the initial one, while the corresponding
solid curve jumps to $M_{\rm BH}\simeq 10^{7}\,M_\odot$ in a
single, effectively instantaneous transition. From that point
onward, the BH grows exclusively through baryonic accretion,
following a smooth Eddington-limited track that eventually
exhausts the baryonic reservoir at low redshift. As $\mu$
decreases, the redshift at which the cloud is absorbed shifts
progressively toward smaller values, occurring at
$z\simeq 34$ for $\mu=10^{-17}\,{\rm eV}$, at $z\simeq 15$ for
$\mu=10^{-18}\,{\rm eV}$, and only around $z\simeq 10$--$11$ for
$\mu=10^{-19}\,{\rm eV}$. In all cases, the scalar phase leaves
the BH at nearly the same post-absorption mass, $\sim
M_{0}+M_{c,0}\simeq 10^{7}\,M_\odot$, and the subsequent
Eddington-driven growth follows a common baryonic asymptote.
Because the scalar reservoir is small in absolute terms, the main
effect of $\mu$ in this regime is therefore \emph{when} the
scalar contribution is delivered, rather than \emph{how much} mass
is transferred; the arrival redshift $z_{\rm reach}(10^{9}
M_\odot)$ is correspondingly set almost entirely by the length of
the baryonic phase that follows the cloud absorption.

\begin{figure}[t]
    \centering
    \includegraphics[width=\columnwidth]{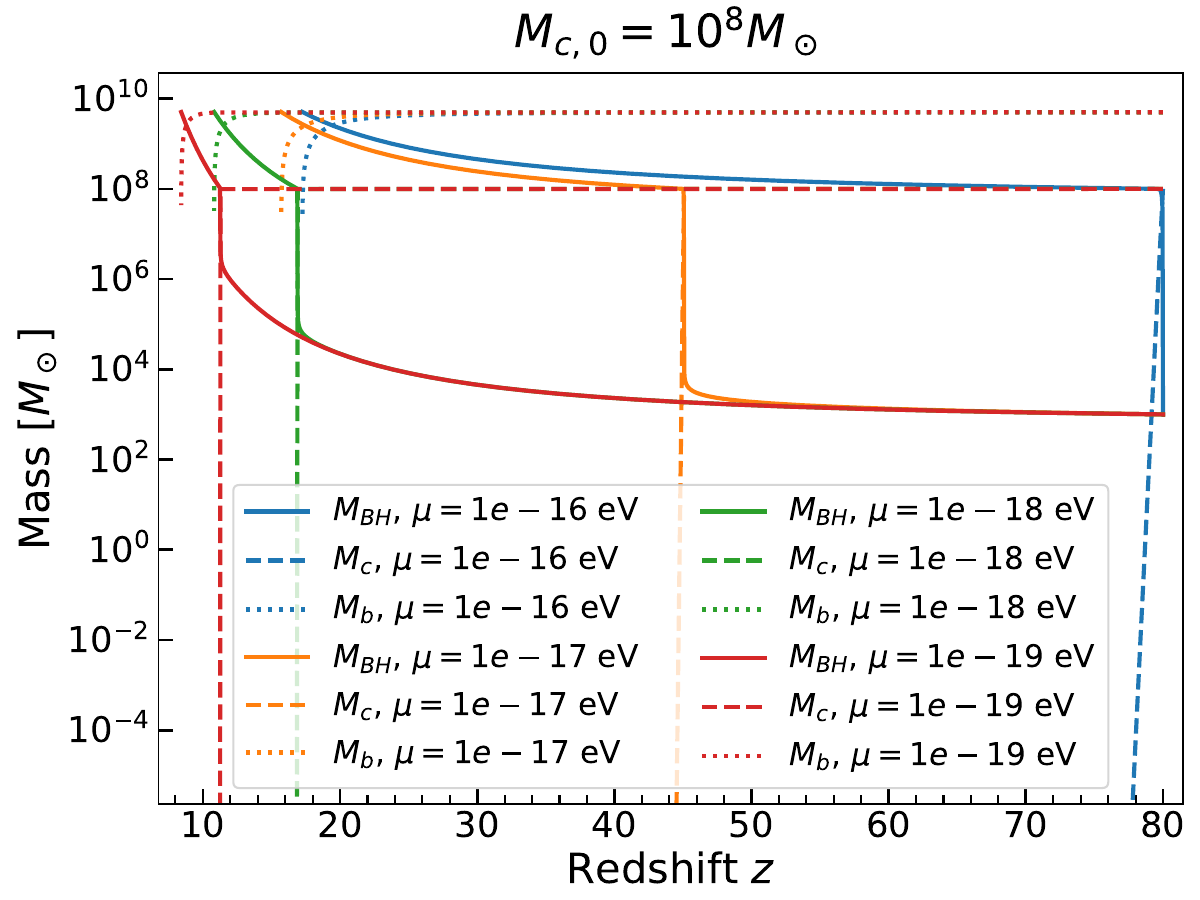}
    \caption{Redshift evolution of $M_{\rm BH}$ (solid), $M_{c}$
    (dashed), and $M_{b}$ (dotted) for $M_{c,0}=10^{8}\,M_\odot$,
    with fixed $M_{0}=10^{3}\,M_\odot$ and
    $M_{b,0}=5\times 10^{9}\,M_\odot$, for four representative
    values of the boson mass $\mu$.}
    \label{fig:evol_mc1e8}
\end{figure}

Figure~\ref{fig:evol_mc1e8} shows the same diagnostic for a
scalar reservoir enhanced by one order of magnitude,
$M_{c,0}=10^{8}\,M_\odot$. The qualitative structure of the
evolution remains the one identified in the previous case, but
the quantitative role of the scalar channel becomes appreciably
more important. The characteristic depletion redshift shifts
slightly toward higher values, at $z\simeq 80$, $z\simeq 45$,
$z\simeq 17$, and $z\simeq 11$ for
$\mu=10^{-16},\,10^{-17},\,10^{-18},\,10^{-19}\,{\rm eV}$
respectively, reflecting the mild sensitivity of the transition
epoch to the amount of scalar matter available. More importantly,
the post-absorption BH mass now reaches
$\sim 10^{8}\,M_\odot$, i.e.\ one order of magnitude closer to
the reference $10^{9}\,M_\odot$ threshold than in
Fig.~\ref{fig:evol_mc1e7}. As a consequence, the subsequent
baryonic phase starts from a substantially more massive seed and
completes the growth to the SMBH regime over a shorter interval
in cosmic time. This behavior confirms, at the level of
individual trajectories, the hierarchical interpretation
introduced in the main text: the boson mass $\mu$ selects when
and how efficiently the scalar channel is activated, whereas the
initial cloud mass $M_{c,0}$ dictates the amplitude of the mass
step imparted to the BH during the scalar-dominated phase. The
baryonic reservoir remains essentially unaffected during the
scalar phase and is depleted only at lower redshift, once the BH
has grown massive enough for the Eddington rate to become
comparable to the reservoir content.

\begin{figure}[t]
    \centering
    \includegraphics[width=\columnwidth]{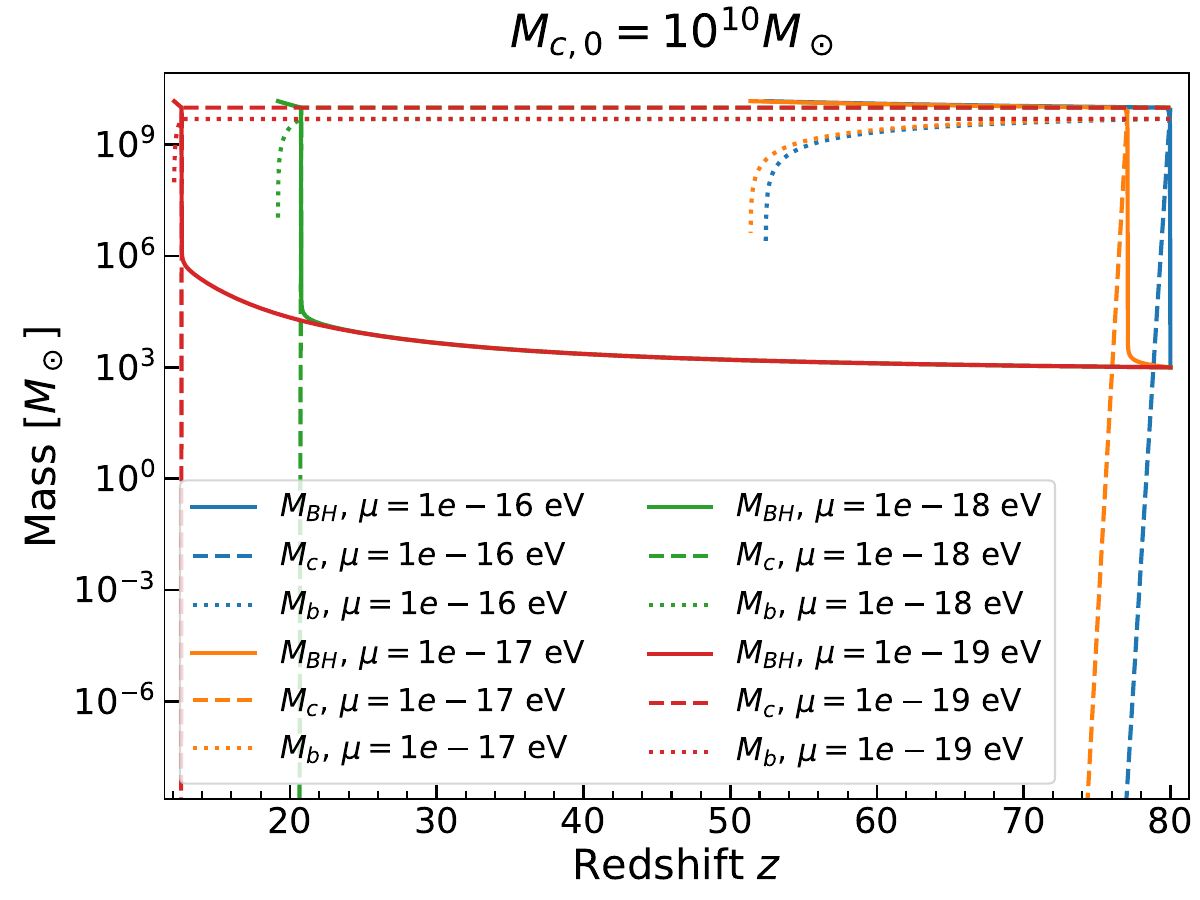}
    \caption{Redshift evolution of $M_{\rm BH}$ (solid), $M_{c}$
    (dashed), and $M_{b}$ (dotted) for
    $M_{c,0}=10^{10}\,M_\odot$, with fixed $M_{0}=10^{3}\,M_\odot$
    and $M_{b,0}=5\times 10^{9}\,M_\odot$, for four representative
    values of the boson mass $\mu$.}
    \label{fig:evol_mc1e10}
\end{figure}

Figure~\ref{fig:evol_mc1e10} corresponds to the largest scalar
reservoir considered in the parameter scan,
$M_{c,0}=10^{10}\,M_\odot$, which exceeds the baryonic reservoir
by a factor of two. This is the regime in which the scalar
channel can most clearly dominate the growth history. For the two
largest boson masses, $\mu=10^{-16}\,{\rm eV}$ and
$\mu=10^{-17}\,{\rm eV}$, the cloud is absorbed almost
immediately, at $z\simeq 78$--$80$, and the BH mass jumps from
$M_{0}=10^{3}\,M_\odot$ to values of order
$M_{\rm BH}\simeq M_{c,0}=10^{10}\,M_\odot$ within a very short
interval of cosmic time. This near-instantaneous transfer of an
extraordinarily large scalar mass has two important consequences.
First, the BH reaches the SMBH regime already at $z\gtrsim 70$,
i.e.\ significantly earlier than in any of the configurations
shown for smaller $M_{c,0}$, which illustrates in the sharpest
possible way the ``dark-to-black'' regime discussed in the main
text. Second, once the BH mass is of order $10^{10}\,M_\odot$,
the Eddington rate becomes correspondingly large, so that the
finite baryonic reservoir is subsequently consumed on a short
timescale: the dotted curves associated with the two highest
values of $\mu$ display a rapid drop at $z\simeq 50$--$55$,
signaling the exhaustion of $M_{b}$ well before the present
epoch. For the two lower boson masses,
$\mu=10^{-18}\,{\rm eV}$ and $\mu=10^{-19}\,{\rm eV}$, the cloud
is instead absorbed at $z\simeq 20$ and $z\simeq 13$
respectively, so that the SMBH-scale jump occurs at
substantially lower redshifts; in these cases the baryonic
reservoir remains essentially intact throughout the integration,
since the BH becomes massive enough to consume it only very close
to the present epoch. Figure~\ref{fig:evol_mc1e10} thus provides
a particularly clean illustration of how, in the regime
$M_{c,0}\gtrsim M_{b,0}$, both the arrival redshift and the
subsequent depletion history of the baryonic reservoir are
essentially controlled by the boson mass, while the scalar
channel dictates the qualitative shape of the growth trajectory.

Taken together, Figs.~\ref{fig:evol_mc1e7}--\ref{fig:evol_mc1e10} complement the reference case $M_{c,0}=10^9M_\odot$ shown in the main text and make two trends particularly clear. At fixed $\mu$, increasing $M_{c,0}$ increases the mass transferred during the scalar-dominated phase and can eventually drive the BH into the SMBH regime without requiring substantial prior baryonic growth. At fixed $M_{c,0}$, changing $\mu$ primarily shifts the epoch at which this transfer occurs, with larger values generally producing earlier cloud depletion. The combined dependence on $\mu$ and $M_{c,0}$ therefore determines both the arrival redshift $z_{\rm reach}(10^9M_\odot)$ and the relative importance of scalar and baryonic accretion. These trajectories reinforce the main conclusion that the dark-to-black regime is not generic, but arises only when the boson mass and the available scalar reservoir jointly place the system in the efficient-transfer region.

\end{document}